\documentclass[twocolumn]{aastex631}

\usepackage{graphicx}
\usepackage{dcolumn}
\usepackage{bm}
\usepackage{amsmath,amssymb,framed,amsthm,xcolor} 
\usepackage{array,multirow,graphicx}
\usepackage{natbib}
\usepackage{float}
\usepackage[makeroom]{cancel}

\begin{document}

\title{Polarized anisotropic synchrotron emission and absorption and its application to Black Hole Imaging}
\author{Alisa Galishnikova}
\altaffiliation[alisag@princeton.edu]{}
\affiliation{Department of Astrophysical Sciences, Princeton University, 4 Ivy Lane, Princeton, NJ 08544, USA}

\author{Alexander Philippov}
\affiliation{Department of Physics, University of Maryland, College Park, MD 20742, USA}

\author{Eliot Quataert}
\affiliation{Department of Astrophysical Sciences, Princeton University, 4 Ivy Lane, Princeton, NJ 08544, USA}

\begin{abstract}
Low-collisionality plasma in a magnetic field generically develops anisotropy in its distribution function with respect to the magnetic field direction. Motivated by the application to radiation from accretion flows and jets, we explore the effect of temperature anisotropy on synchrotron emission. We derive analytically and provide numerical fits for the polarized synchrotron emission and absorption coefficients for a relativistic bi-Maxwellian plasma (we do not consider Faraday conversion/rotation). Temperature anisotropy can significantly change how the synchrotron emission and absorption coefficients depend on observing angle with respect to the magnetic field. The emitted linear polarization fraction does not depend strongly on anisotropy, while the emitted circular polarization does. We apply our results to black hole imaging of Sgr A* and M87* by ray-tracing a GRMHD simulation and assuming that the plasma temperature anisotropy is set by the thresholds of kinetic-scale anisotropy-driven instabilities. We find that the azimuthal asymmetry of the 230 GHz images can change by up to a factor of $3$, accentuating ($T_\perp > T_\parallel$) or counteracting ($T_\perp < T_\parallel$) the image asymmetry produced by Doppler beaming. This can change the physical inferences from observations relative to models with an isotropic distribution function, e.g., by allowing for larger inclination between the line of sight and spin direction in Sgr A*. The observed image diameter and the size of the black hole shadow can also vary significantly due to plasma temperature anisotropy. We describe how the anisotropy of the plasma can affect future multi-frequency and photon ring observations. In Appendices we calculate kinetic anisotropy-driven instabilities (mirror, whistler, and firehose) for relativistically hot plasmas.
\end{abstract}

\section{Introduction} \label{sec:intro}

Synchrotron emission produced by relativistic electrons in the presence of a magnetic field appears in many astrophysical systems. It is the source of emission across much of the electromagnetic spectrum in pulsar wind nebulae and jets from neutron stars and black holes (BHs). Synchrotron emission is also the source of the mm-wavelength radio emission observed on event-horizon scales in M87* and Sgr A* by the Event Horizon Telescope (EHT) \citep{EHT2019ApJ1,EHT2022ApJ}. 

Models of synchrotron emission from astrophysical plasmas typically assume that the plasma has a thermal or power-law distribution function or a hybrid of the two, such as a kappa distribution function.  The latter two are motivated by the power-law (non-thermal) synchrotron spectra often observed from astrophysical sources.   Another explicit assumption typically made is that the electron distribution function is isotropic relative to the local magnetic field, i.e., that the electrons have the same temperature or energy density in all directions.\footnote{An exception to this is in very strongly magnetized plasmas such as neutron star magnetospheres where the synchrotron cooling time is so short that the perpendicular energy is nearly instantaneously radiated away. In this paper we are focused on applications with weaker magnetic fields, such as black hole accretion flows and jets.}

In the presence of dynamically strong magnetic fields, the assumption of an isotropic electron distribution function is not theoretically or observationally well-motivated. By dynamically strong here, we mean an energy density in the magnetic field similar to or larger than that in the plasma. Such magnetized collisionless (and weakly collisional) plasmas can readily depart from thermal equilibrium and develop anisotropies with respect to the local magnetic field direction \citep{Quataert_2002}. Although the distribution function will in general be gyrotropic (isotropic in the plane perpendicular to the magnetic field), it can have significant anisotropies parallel and perpendicular to the local magnetic field \citep{1983hppv}.

There is extensive observational evidence for such anisotropy in the solar corona and solar wind \citep{SolarWind2009}. In the most extreme cases, Oxygen ions in the solar corona have perpendicular temperatures that are a factor of $\sim 10-100$ times that of their parallel temperature \citep{Cranmer1999}. This anisotropy is in fact critical to interpreting spectroscopy of the solar corona. By analogy, one might expect that anisotropy in the electron distribution function could be important for interpreting synchrotron radiation from astrophysical plasmas. This is particularly true in high spatial resolution observations where our viewing angle relative to the local magnetic field likely changes significantly across the image (e.g., the EHT or radio interferometry more generally).

The anisotropy in a plasma's distribution function cannot, however, grow without bound. It is limited by kinetic-scale instabilities such as the mirror, whistler, firehose, and ion cyclotron instabilities \citep{Rosenbluth1956, SouthwoodKivelson93, ChandrasekharKaufmanWatson1958,RudakovSagdeev1961, NonrelWhistlerSudan, Gary1992}. When the anisotropy in the distribution function becomes too large (relative to the threshold of the instability\footnote{Some instabilities, e.g., the ion cyclotron instability, formally do not have a threshold, but their growth rate becomes sufficiently small at low anisotropies that in practice they do.}), such instabilities rapidly grow, driving the anisotropy towards the instability threshold. This endows the plasma with an effective collisionality that acts to partially isotropize the distribution function.  A very rough rule of thumb is that instabilities set in vigorously when the fractional temperature anisotropy satisfies $|\Delta T/T| \gtrsim \mathcal{O}(\beta^{-1})$ (where $\Delta T$ is the temperature anisotropy and $\beta$ is the ratio of thermal to magnetic energy).  Anisotropy can thus be much larger in strongly magnetized plasmas with $\beta \lesssim 1$. Anisotropy in the distribution function is thus expected to be particularly important in jets and in models of accretion flows with dynamically strong magnetic fields, such as the Magnetically Arrested Disc (MAD) models favored by EHT observations of M87* \citep{EHT2021ApJmagneticField}.

Observations of protons and electrons in the solar wind show that they obey the expected anisotropy-driven instability thresholds and that the anisotropy is larger at lower $\beta$ \citep{SolarWind2009} (however, the measured anisotropy is smaller than the instability thresholds at $\beta \lesssim0.1$).  We expect that in accretion flows and jets, inflow, outflow, and heating of the plasma will likewise drive temperature anisotropies to the point that instabilities set in \citep{Foucart2017MNRAS}. Global axisymmetric GR kinetic simulations of collisionless plasma accreting onto a BH indeed find the growth of the mirror and firehose instability and that they regulate the plasmas's temperature anisotropy \citep{GRPIC2023PhRvL}. 

Motivated by the potential importance of an anisotropic distribution function in synchrotron emitting plasmas, in this paper we theoretically calculate emission and absorption of polarized synchrotron radiation for a physically motivated gyrotropic distribution function. The study of polarized synchrotron radiation dates back to the work of \citet{Westfold1959ApJ}, who studied emission from an ultra-relativistically gyrating electron. General formulae for Stokes parameters for ultra-relativistic synchrotron emission from an assemble of electrons can be found in the review of \citet{GinzburgSyrovatskii1965}, who noted that a substantial amount of circular polarization is present only in the case of a highly anisotropic pitch-angle distribution. \citet{Melrose1971} presented the general equations for Stokes parameters for an arbitrary anisotropic distribution function separable in momentum and pitch-angle, while \citet{Sazonov1972} focused on the case of a power-law momentum distribution with a separable pitch-angle anisotropy.

In the last few decades, the study of synchrotron radiation was extended to a broader range of validity and a number of different distribution functions via numerical integration methods \citep{Mahadevan1996ApJ, Shcherbakov2008ApJ, Leung2011ApJ, Pandya2016ApJ, Pandya2018ApJ, Dexter2016MNRAS}. This is useful for improving analytical results at arbitrary frequency, emission direction with respect to the magnetic field, and distribution function.  These works provide fits for the Stokes emissivities, absorptivities, and rotativities that have been widely used in modeling polarized synchrotron radiation from accreting black holes, particularly in the context of the EHT sources M87* and Sgr A* [e.g., \citet{Dexter2016MNRAS, ipole2018MNRAS, blacklight2022ApJS} and others; see also \citet{GRRTcomparison2020ApJ}]. However, no pitch-angle anisotropy was considered in these studies.

In this paper, we extend previous work on synchrotron radiation by studying the intrinsic emission from an ensemble of electrons with an anisotropic relativistic distribution function. We focus on the case of a relativistic generalization of a bi-Maxwellian that has different temperatures perpendicular and parallel to the local magnetic field (\S \ref{sec:expressions}) and provide fits for the polarized emissivity and absorption coefficients in Section~\ref{sec:fits}. We defer the case of an anisotropic power-law distribution function to future work.   We also defer the calculation of Faraday rotation and conversion coefficients for an anisotropic distribution function to future work. We then implement these expressions in a GR radiative transfer code to ray trace GRMHD MAD simulations and study the impact of pitch-angle anisotropy on the observable quantities (Section~\ref{sec:imaging}). Finally, in Section~\ref{sec:sum} we summarise the application of our results to current and future EHT observations.  

\section{Synchrotron Emission From Gyrotropic Distribution Functions} \label{sec:expressions}

In this section, we describe radiation transfer and emission produced by electrons with a gyrotropic distribution function $f(\gamma, \xi)$ in the presence of a background magnetic field ${\bf B}$; $\gamma$ and $\xi$ denote the Lorentz factor of electrons and pitch angle with the magnetic field respectively;  we will use $\mu = \cos \xi$ and $\xi$ interchangeably in what follows.  Throughout the paper, $m_e$, $e$, and $c$ are constants that stand for electron mass, electron charge, and speed of light. Therefore, the momentum of a particle with velocity ${\bf v}$ is ${\bf p}=m_e \gamma {\bf v}$ and $\beta={\bf v} /c$. In what follows, we normalize the frequency of emission $\nu$ by a non-relativistic cyclotron frequency given by $\nu_c = eB/2 \pi m_e c$. The angle between the propagation direction along the wavevector ${\bf k}$ and the background magnetic field ${\bf B}$ is set by $\theta_B$.

Polarized emission is described in the Stokes basis as $I_a=\{ I,Q,U,V \} ^T$, where $I$ stands for intensity, $Q$ and $U$ describe linear polarization, and $V$ describes circular polarization. Given emissivities $j_a=\{j_I,j_Q,j_U,j_V\}^T$, absorption coefficients $\alpha_a=\{\alpha_I,\alpha_Q,\alpha_U,\alpha_V\}^T$, and Faraday rotativities $\rho_a=\{\rho_Q, \rho_U, \rho_V\}^T$, the polarized emission can then be found using [see, e.g, \citet{Leung2011ApJ}]
\begin{equation}
    \frac{d I_a}{d s} = j_a - M_{ab} I_b,
\end{equation}
where $M_{ab}$ is the Mueller matrix,
\begin{equation}
M_{ab} = 
    \begin{pmatrix}
    \alpha_I & \alpha_Q & \alpha_U & \alpha_V \\
    \alpha_Q & \alpha_I & \rho_V & -\rho_U \\
    \alpha_U & -\rho_V & \alpha_I & \rho_Q \\
    \alpha_V & \rho_U & -\rho_Q & \alpha_I 
    \end{pmatrix},
\end{equation}
where $U$ components vanish if ${\bf B}$ is aligned with $U$: $j_U=0$, $\alpha_U=0$, and $\rho_U=0$. Then $I$ components of $j_a$ and $\alpha_a$ describe total emission, $Q$ components describe linearly polarised emission and $V$ describe circularly polarised emission, while $\rho_Q$ and $\rho_V$ account for Faraday conversion and rotation respectively. In this work, we focus on emissivities $j_a$ and absorption coefficients $\alpha_a$, while Faraday rotativities $\rho_a$ will be studied in future work.

We need to evaluate $j_a$, $\alpha_a$, and $\rho_a$ through $f(\gamma, \xi)$ to describe the radiation emission and transfer. In the Stokes basis at frequency $\nu$ [see, e.g., \citet{Leung2011ApJ}]:
\begin{equation}
    \begin{split}
    j_a &= \frac{2 \pi e^2 \nu^2}{c} \int d^3 p f(\gamma,\xi) \sum_{n=1}^\infty\delta (y_n) K_a(z),\\
    \alpha_a &= \frac{2 \pi \nu }{m_ec^2} \int d^3 p Df(\gamma,\xi) \sum_{n=1}^\infty\delta (y_n) K_a(z),
    \end{split}
\label{eq:S_equations}
\end{equation}
where $\delta(y_n)$ is a delta function of argument $y_n=n \nu_c / \gamma - \nu(1-\beta \cos \xi \cos \theta_B)$, $z = \nu \gamma \beta \sin \theta_B \sin \xi / \nu_c$, $d^3p=2\pi m_e^3 c^3 \gamma^2 \beta d \gamma d \cos \xi$ for a gyrotropic $f(\gamma, \xi)$, and
$Df$ is an operator that includes a full derivative of the distribution function:
\begin{equation}
\begin{split}
   & Df  \equiv \left( k_\parallel \frac{\partial}{\partial p_\parallel } + \frac{\omega - k_\parallel v_\parallel}{v_\perp} \frac{\partial}{\partial p_\perp} \right) f(\gamma, \xi) \\
     &= \frac{2 \pi \nu}{m_e c^2} \left( \frac{\partial}{\partial \gamma} + \frac{\beta \cos \theta_B - \cos \xi}{\beta^2 \gamma} \frac{\partial}{\partial \cos \xi} \right) f(\gamma, \xi),
\end{split}
\label{eq:Df} 
\end{equation}

In Equation \ref{eq:S_equations}, $K_a$ is defined as
\begin{equation} 
K_a=
\begin{cases}
M^2 J_n^2(z) + N^2 J'^2_n(z), a=I,\\
M^2 J_n^2 (z) - N^2 J'^2_n(z), a=Q,\\
0,  a=U,\\
2MNJ_n(z) J'_n(z), a=V,
\end{cases}\label{eq:kernel}
\end{equation}
where $J_n$ is a Bessel function of the first kind, $M=(\cos \theta_B - \beta \cos \xi)/\sin \theta_B$, and $N=\beta \sin \xi$. Given $f(\gamma, \xi)$, one can find $j_a$ and $\alpha_a$ through Equations~\ref{eq:S_equations}, \ref{eq:Df}, and \ref{eq:kernel}. 

\subsection{Anisotropic electron distribution function}
We will use an anisotropic distribution function $f(\gamma, \xi)$ for emitting electrons, written in cgs units:
\begin{equation}
\begin{split}
& f(p_\perp, p_\parallel )  =   \frac{n_e \eta^{1/2}}{4 \pi m_e^3 c^3 \epsilon_\perp K_2 (1/\epsilon_\perp)} \times \\
& \exp{(- \sqrt{1 + (p_\perp / m_ec)^2 + \eta (p_\parallel/m_ec)^2} / \epsilon_\perp)},
\end{split}
\end{equation}
where  $n_e$ is the electron number density, $ \epsilon_\perp = kT_{\perp,e}/m_e c^2$ is the dimensionless perpendicular electron temperature, $K_2$ is the modified Bessel function of the second kind, $p_\perp$ and $p_\parallel$ stand for the relativistic momentum perpendicular and parallel to the magnetic field direction.  Here $\eta$ is a measure of anisotropy, with $\eta =1$ corresponding to an isotropic relativistic Maxwellian distribution function. In the non-relativistic limit, $T_{\perp,e}/ T_{\parallel,e} = \eta$, while $T_{\perp,e}/ T_{\parallel,e} \approx \eta^{0.8}$ in the ultra-relativistic limit (see Appendix \ref{ap:model} for a detailed fit). Transforming $f(p_\perp, p_\parallel)$ to $\gamma-\xi$ variables:
\begin{equation}
\begin{split}
& f(\gamma, \xi ) =   \frac{n_e \eta^{1/2}}{4 \pi m_e^3 c^3 \epsilon_\perp K_2 (1/\epsilon_\perp)} \times \\
& \exp{(- \sqrt{1 + (\gamma^2 - 1)(\sin^2 \xi + \eta \cos^2 \xi)} / \epsilon_\perp)}.
\end{split}\label{eq:df}
\end{equation}
In the limit of high $\gamma$:
\begin{equation}
\begin{split}
f(\gamma, \xi) & =   \frac{n_e \eta^{1/2}}{4 \pi m_e^3 c^3 \epsilon_\perp K_2 (1/\epsilon_\perp)} \times \\
& \exp{(- \gamma \sqrt{1 + (\eta - 1) \cos^2 \xi} / \epsilon_\perp)} \\
& = \frac{n_e \eta^{1/2}}{4 \pi m^3 c^3 \epsilon_\perp K_2 (1/\epsilon_\perp)} \times \exp{(- \gamma / \epsilon^\ast _\perp)},
\end{split}
\label{eq:df_high_gamma}
\end{equation}
where 
\begin{equation}
\epsilon^\ast_\perp=\epsilon^\ast_\perp(\xi) = \frac{\epsilon_\perp}{\sqrt{1 + (\eta - 1) \cos^2 \xi}}
\label{eq:epsperpstar}
\end{equation}
is the new renormalized temperature.  Thus, in the high-$\gamma$ limit, the temperature in the distribution function depends on both the anisotropy $\eta$ and pitch angle or $\mu = \cos \xi$. In the isotropic case $\eta=1$, the temperature is described by $\epsilon = \epsilon_\perp = \epsilon_\parallel$ in all directions. Note that in the analytical fitting functions in the next subsection, $\epsilon^\ast_\perp$ will be evaluated at $\xi = \theta_B$ because the radiation is beamed along the local direction of motion of relativistic electrons (as is standard in synchrotron radiation, see Appendix~\ref{ap:comparison} for details). In our numerical evaluations, however, we integrate and sum over $\xi$ and $\theta_B$ separately using Equation~\ref{eq:S_equations}.

The total derivative $Df$ that is used in calculating $\alpha_a$ in Equation~\ref{eq:S_equations} contains
\begin{equation}
\begin{split}
\partial_\gamma f(\gamma, \xi) & = -  \frac{\gamma}{\epsilon_\perp} \frac{1 + \mu^2 (\eta-1)}{\sqrt{\gamma^2 + (\gamma^2 - 1)(\eta-1)\mu^2}} f(\gamma, \xi), \\
\partial_\mu f(\gamma, \xi) & =  - \frac{(\gamma^2-1)(\eta-1)\mu}{\epsilon_\perp \sqrt{\gamma^2 + (\gamma^2 - 1)(\eta-1)\mu^2}} f(\gamma, \xi).\\
\end{split}
\end{equation}
While $\partial_\mu f(\gamma, \xi)$ is non-zero, we find that the absorption coefficients change negligibly if we include this term. This is due to the prefactor it goes with in equation \ref{eq:Df} since $\gamma \gg 1$ and the absorption is mainly concentrated around $\xi \approx \theta_B$ (see Appendix \ref{ap:comparison} for details). 

\subsection{Emissivities and absorption coefficients}\label{sec:fits}
\begin{figure}
    \centering
    \includegraphics[width=\columnwidth]{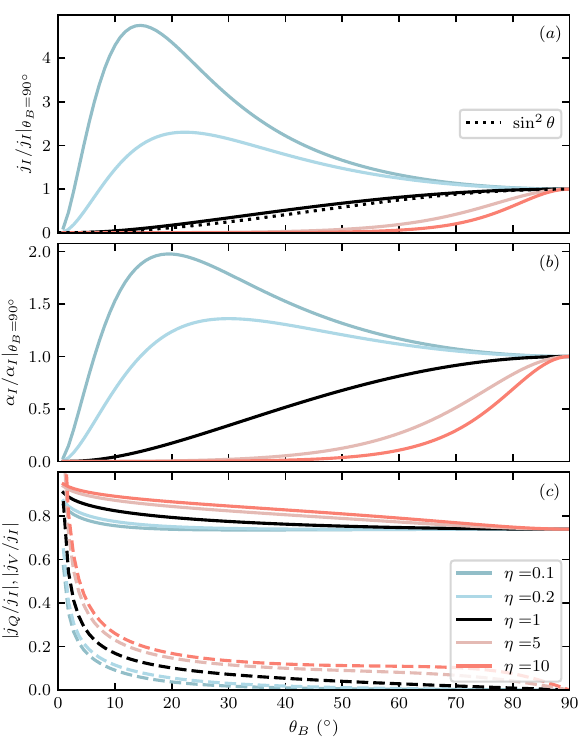}
    \caption{Emissivity $j_{I}$ (a), absorption $\alpha_{I}$ (b), and emitted polarization fractions (c) as functions of the angle between propagation direction and magnetic field $\theta_B$ at different anisotropy values: $T_\perp < T_\parallel$ ($\eta<1$, blue), isotropic ($\eta=1$, black), and $T_\perp > T_\parallel$ ($\eta>1$, red). The free parameters are $\nu/\nu_c=10^3$ and $\epsilon_\perp=10$ (near the peak of the optically thin synchrotron spectrum for an isotropic distribution function). In (c) the emitted linear $|j_{Q}|/j_{I}$ and circular $|j_{V}|/j_{I}$ polarization fractions are shown by solid and dashed lines respectively. In (a) $\sin^2 \theta_B$ dependence is shown by a black dotted line.}
    \label{fig:jnu}
\end{figure}

We obtain the following fits for emissivities and absorption coefficients for a relativistic plasma with an anisotropic bi-Maxwellian distribution function (see Appendix \ref{ap:comparison} for details on the derivation):
\begin{equation}
j_a = 
\begin{cases}
\eta^{1/2} \mathcal{K}_a(\epsilon^\ast_\perp,\epsilon_\perp) j_{a, {\rm iso}}(\epsilon=\epsilon_\perp^\ast, \nu/\nu_c, \theta_B), a = \{ I, Q\}\\
\eta^{3/2}\mathcal{K}_a(\epsilon^\ast_\perp,\epsilon_\perp) j_{a, {\rm iso}}(\epsilon=\epsilon_\perp^\ast, \nu/\nu_c, \theta_B), a = V,
\end{cases}\label{eq:fits}
\end{equation}
where
\begin{equation}
\mathcal{K}_a(\epsilon^\ast_\perp/\epsilon_\perp) = 
\begin{cases}
(\epsilon^\ast_\perp/ \epsilon_\perp) [K_2(1/\epsilon^\ast_\perp)/ K_2(1/\epsilon_\perp)], a = \{ I, Q\}\\
(\epsilon^\ast_\perp/ \epsilon_\perp)^3 [K_2(1/\epsilon^\ast_\perp)/ K_2(1/\epsilon_\perp)], a = V,
\end{cases}\label{eq:fits_factor}
\end{equation}
and $K_2(1/\epsilon_\perp) \approx 2 \epsilon_\perp^2$ when $\epsilon_\perp \gg 1$. Here $\epsilon^\ast_\perp$ is evaluated at $\xi = \theta_B$ and 
$j_{a, {\rm iso}}(\epsilon=\epsilon_\perp^\ast, \nu/\nu_c, \theta_B)$ and $\alpha_{a, {\rm iso}}(\epsilon=\epsilon_\perp^\ast, \nu/\nu_c, \theta_B)$ correspond to emission and absorption in the case of an isotropic relativistic Maxwellian. Absorption coefficients $\alpha_a$ can be obtained via Kirchoff's law for a thermal distribution function:
\begin{align} 
j_{a, {\rm iso}} - \alpha_{a, {\rm iso}} B_\nu=0 ,
\end{align}
where $B_\nu(T_{\perp,e})=\frac{2h\nu^3}{c^2} (\exp{(h\nu/kT_{\perp,e})-1})^{-1}$ for an anisotropic distribution has the same functional form as in the isotropic case (with $T_{\perp,e}=T_e$ for an isotropic distribution). Equations \ref{eq:fits} correspond to total intensity and linearly polarized emissivities with the same functional form as in an isotropic plasma but with a temperature $\epsilon_\perp^\star$ that depends on the observer-angle $\theta_B$ due to the anisotropy in the distribution function relative to the magnetic field (the factors of $\eta^{1/2} \mathcal{K}_a(\epsilon^\ast_\perp,\epsilon_\perp) \approx \eta^{1/2} (\epsilon^\star_\perp/\epsilon_\perp)^3$ in Equation \ref{eq:fits} reflect the change in normalization of the distribution function due to the different number of particles whose radiation is beamed in the direction of the observer). By contrast, the Stokes V (circular polarization) emissivity in Equation \ref{eq:fits} differs by a larger factor because of a change in the efficiency of producing circularly polarized radiation for an anisotropic distribution function (see Eq. \ref{eq:jtensor} and \ref{eq:jV} in Appendix \ref{ap:comparison}).

Given Equation \ref{eq:fits}, it is straightforward to derive the location of the peak of the optically thin emission $\nu j_I(\epsilon, \nu, \theta_B, \eta)$ as
\begin{equation}
    \nu_{\rm peak} \approx 36.7 \nu_c \epsilon_\perp^{\ast 2} \sin \theta_B \simeq \frac{ 36.7 \nu_c \epsilon_\perp^2 \sin \theta_B}{1 + (\eta - 1) \cos^2 \theta_B},
    \label{eq:nu_peak}
\end{equation}
which shifts to lower (higher) frequencies with increasing (decreasing) $\eta$ at a fixed $\epsilon_\perp$ and $\theta_B$ (though we show below in Figure \ref{fig:jnu} that $\eta$ changes the efficiency of producing radiation as a function of $\theta_B$).

We find good agreement between our analytic expressions and numerical calculation over a wide parameter range using a publicly available code {\tt symphony}, which integrates Equations \ref{eq:S_equations} for a given distribution function $f(\gamma,\xi)$. For a detailed derivation of the fits given by Equations \ref{eq:fits}, full expressions for $j_{a,{\rm iso}}$ and $\alpha_{a,{\rm iso}}$, and the comparison with numerical solutions see Appendix \ref{ap:comparison}. The fits presented here become inaccurate for $\epsilon^\star_\perp \lesssim 3$ and low frequencies $\nu/\nu_c \lesssim 10$, where the isotropic fits that we scale to in Equation \ref{eq:fits} themselves become inaccurate.

We demonstrate the resulting emission properties in Figure \ref{fig:jnu}, where $j_{I}$ (a), $\alpha_{I}$ (b), and the emitted linear and circular polarization fractions (solid and dashed lines in (c) respectively) are shown as functions of the angle between propagation direction and magnetic field $\theta_B$ at different values of anisotropy parameter $\eta$, represented by different colors (we intentionally choose relatively large anisotropy to highlight the large differences in synchrotron radiation possible in this limit). The parameters used in this Figure are a high frequency of $\nu/\nu_c=10^3$ and temperature of $\epsilon_\perp=10$. The isotropic case (black line) closely follows a $\sin^2 \theta_B$ dependence of $j_I$ (dotted) in panel (a) due to the frequency being near the peak of the synchrotron emissivity.

Figure \ref{fig:jnu} shows that there are significant differences in the synchrotron emission/absorption for an anisotropic plasma distribution, compared to the isotropic case. This change can be understood as a renormalization of the number of relativistic particles emitting toward the observer at $\theta_B$. In particular, the plasma is less prone to emitting along the magnetic field at $\eta > 1$ ($T_\perp > T_\parallel$, red lines), hence the rapid fall off of $j_{\nu, I}$ with decreasing $\theta_B$, compared to $\eta \equiv 1$. That is, the emission is even more concentrated towards $\theta_B=90^\circ$ when $\eta > 1$. For the opposite anisotropy, $\eta<1$ and $T_\parallel > T_\perp$ (blue lines), the number of particles capable of emitting along the magnetic field direction increases. Thus, more emission can be produced 
at smaller $\theta_B$ (along the magnetic field), relative to the isotropic case with $\eta\equiv1$. The unpolarised absorption coefficient $\alpha_I$ (b) shows a similar but smaller dependence on $\eta$ as $j_I$.

The polarization fractions have a weaker dependence on $\eta$. This is shown in Figure \ref{fig:jnu} (c) with solid and dashed lines for the intrinsic linear $|j_Q|/j_I$ and circular $|j_V|/j_I$ polarization fractions, respectively.  

Quantitatively, both $|j_Q|/j_I$ and $|j_V|/j_I$ are higher for higher $\eta$ but the change is particularly modest for the intrinsic linear polarization $|j_Q|/j_I$. Since most of the emission comes from small (large) angles for $\eta<1$ ($\eta>1$), the emitted circular polarisation degree can significantly vary with $\eta$ due to the change in which pitch angles dominate the emission. In particular, $\eta>1$ is significantly more circularly polarised, and $\eta<1$ is less circularly polarized, compared to emission from an isotropic plasma. This is because $\eta>1$ decreases the effective temperature $\epsilon_\perp^\ast$ by suppressing the parallel temperature at a fixed $\epsilon_\perp$.

\begin{figure*}
    \centering
    \includegraphics[width=\textwidth]{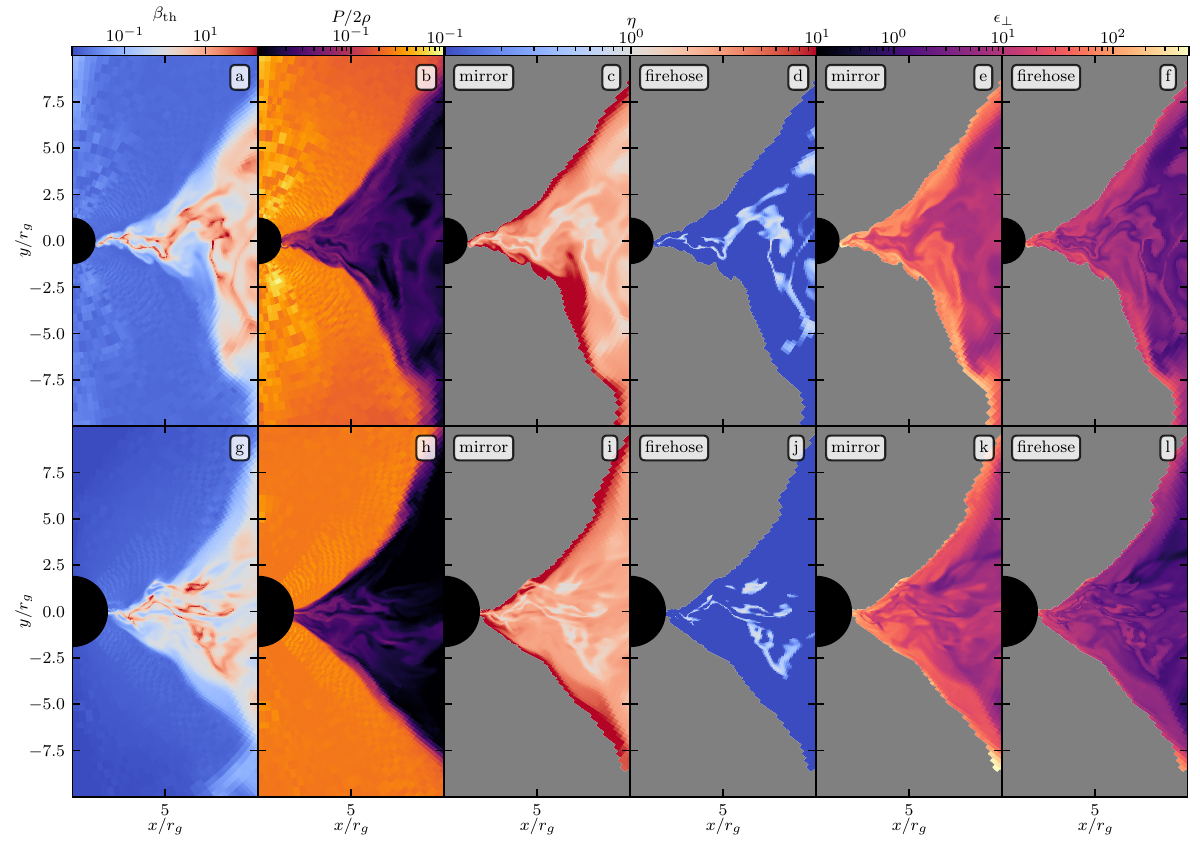}
    \caption{Plasma-$\beta_{\rm th}$ (column 1, a and g), plasma temperature $P/2 \rho$ (column 2, b and h), anisotropy $\eta$ (columns 3 and 4 for mirror and firehose respectively), and normalized electron perpendicular temperature $\epsilon_\perp$ (columns 5 and 6 for mirror and firehose respectively) for $a=0.98$ (at time of $14300r_g/c$, top row) and $a=0.5$ (at time of $16000 r_g/c$, bottom row).}
    \label{fig:epsilon}
\end{figure*}

\section{Black Hole Imaging} \label{sec:imaging}
In this section we study the observational implications of synchrotron emission by a plasma with anisotropic temperatures in the context of black hole accretion flows. Specifically, we focus on the application to the EHT targets Sgr A* and M87* \citep{EHT2019ApJ1, EHT2022ApJ}. Our goal in this initial study is to determine the rough magnitude of the effect and which observables are most sensitive to electron temperature anisotropy. The exact electron temperature anisotropy in the near-horizon plasma is uncertain so we will use general stability arguments to bound the anisotropy and thus the effect of anisotropy on the synchrotron radiation.   

\subsection{Method}
We use a publicly available radiative transfer code {\tt blacklight} \citep{blacklight2022ApJS} to ray trace synchrotron emission in GRMHD simulations and study the resulting intensity and polarization images. We implement the formulas for the emissivity and absorption coefficients of hot electrons with an anisotropic distribution function discussed in \S \ref{sec:expressions} (we use the limit of high temperature such that $K_2(1/x) \approx 2x^2$). Since EHT observational constraints favor highly magnetized models \citep{EHT2021ApJmagneticField}, we restrict our study to a MAD simulation of plasma accreting onto a spinning BH with dimensionless spin parameters of $a=0.98$ and $0.5$. Our results are averaged over $100$ snapshots which span a time of $1000r_g/c$ when the accretion rate and magnetic flux on the horizon are in approximate steady-state (see Appendix \ref{ap:simulations} for details on the simulation setup and choice of this time period). Since the MHD method cannot handle vacuum, our GRMHD simulations have a ceiling plasma magnetization parameter of $\sigma=B^2/[ 4 \pi \rho c]=100$, and we ignore emission from $\sigma > 10$ regions.\footnote{The magnetization in the jet region can, in reality, be significantly larger than the ceiling value set in our GRMHD simulations. These low-density regions with $\sigma \gtrsim 10$ are, however, not expected to contribute significantly to the observed flux at $230$ GHz.}  We choose a BH mass and distance to the BH to match M87*, $M_{BH} = 6.5\times 10^9 M_{\odot}$ and $d = 1.67 \times 10^7$pc, {unless otherwise specified}. In the GRMHD simulations, the plasma number density normalization is a free parameter, which we choose such that the total flux of the image $F_{\nu}$ at $230$ GHz matches EHT observations of M87*, i.e., $0.66$ Jy. The raytraced images have a resolution of $128 \times 128$ cells, with a point camera located at $100r_g$ and inclination [observing angle] of $\theta$. We consider both $\theta = 163 \deg$, appropriate for M87*, as well as less face-on viewing angles to demonstrate the change with viewing angle.    

Since the GRMHD equations evolve a single fluid, while in the plasmas of interest the electrons and ions likely have different temperatures, we have the freedom to set the electron temperature. The heating of collisionless electrons should depend on local plasma parameters, in particular, the magnetic field strength \citep{QuataertGruzinov1999ApJ} via $\beta_{\rm th} = P_{\rm th}/P_{B}$ -- the ratio of thermal pressure to magnetic pressure. To parameterize the electron temperate, we use the widely-employed $R_{\rm high}-R_{\rm low}$ model \citep{RhighRlow2016A&A}. In this model, the ion-to-electron temperature is set by
\begin{equation}
    R = \frac{T_i}{T_e} = \frac{\beta_{\rm th}^2 R_{\rm high} + R_{\rm low}}{1+\beta_{\rm th}^2},
    \label{eq:Rmodel}
\end{equation}
where $\beta_{\rm th}=P_{{\rm th}}/P_{B}$ is the plasma $\beta_{\rm th}$ for an MHD fluid, $R_{\rm high}$ and $R_{\rm low}$ are ion-to-electron temperature ratios in the high and low-$\beta_{\rm th}$ regions respectively. The fluid GRMHD temperature is $T=(T_i + T_e)/2$, and thus $T_e = 2 T/ (1+R)$. In this work, we explore three cases: $R_{\rm high}=1$, $10$ and $100$, while $R_{\rm low}$ is set to $1$ always.

To study the effect of the anisotropy of the plasma on images, we also have the freedom to set the anisotropy parameter $\eta \sim T_{\perp,e}/T_{\parallel,e}$ since the GRMHD simulations have no information about plasma anisotropy. The anisotropy of the plasma is limited by kinetic-scale instability thresholds, which allow for a large anisotropy in low-$\beta_{\rm th}$ regions.  Ion-scale mirror and firehose instabilities are clearly present in global GR kinetic simulations of collisionless plasma accreting onto a BH \citep{GRPIC2023PhRvL} and in kinetic shearing box simulations \citep{Kunz2014PhRvL, Riquelme2015ApJ}. The electrons also contribute to driving mirror, firehose, and whistler instabilities, which are important for setting the electron temperature anisotropy. Since the magnitude of the electron temperature anisotropy in the near-horizon environment is not fully understood, we consider all three limiting cases -- where the plasma sits at the mirror ($\eta>1$), whistler ($\eta>1$), or firehose ($\eta<1$) instability thresholds everywhere.  We then compare these limiting cases to the usually considered isotropic plasma distribution. This should bracket the magnitude of the effect introduced by an anisotropic electron distribution function. We note that in the single-fluid global ``extended GRMHD'' simulations of \citet{Foucart2017MNRAS} in which the pressure anisotropy is a dynamical variable, most of the plasma was near the mirror threshold. If generically true, and applicable to electrons, this would suggest that the mirror and whistler instability thresholds are the most important. 

The microinstability thresholds can be expressed as $T_{\perp,e}/T_{\parallel,e}=g(\beta_e)$, where $g(\beta_e)$ is a function of election plasma$-\beta$, different for each of the anisotropy-driven instability [see Appendix \ref{ap:instability} for a derivation of relativistic mirror, firehose, and whistler instabilities and Appendix \ref{ap:model} for additional details on $g(\beta_e)$ for each instability]. Therefore, our procedure for obtaining $T_{\perp,e}$ and $\eta$ for each of the four instability cases (here and after mirror, whistler, isotropic, firehose) is as follows. We first compute electron-$\beta$ as $\beta_e = 2 \beta_{\rm th}/(1 + R)$, according to Eq.~\ref{eq:Rmodel}. Knowing $\beta_e$ allows us to calculate $T_{\perp,e}/T_{\parallel,e}=g(\beta_e)$ for each case of interest and thus $\eta$ (we consider the relativistic limit, where $T_{\perp, e}/T_{\parallel,e}\approx\eta^{0.8}$).  The perpendicular temperature can then be separately determined from the definition $T_e = (T_{\parallel,e} + 2T_{\perp,e})/3 = T_{\perp,e} (\eta^{1/0.8} + 2)/3 = T_i/R$. Now that we have $\eta$ and $\epsilon_\perp$ for electrons, we can then calculate $j_a$ and $\alpha_a$, given by Equations \ref{eq:fits} (with $\epsilon_{\perp} = kT_{\perp,e}/m_ec^2$).

In Figure~\ref{fig:epsilon} we show an example of the inferred physical conditions in MAD accretion flows from GRMHD simulations: MHD-$\beta_{\rm th}$ (first column), plasma temperature $P/2\rho$ (second column), and the resulting $\eta$ (third and fourth columns for mirror and firehose respectively) and electron $\epsilon_\perp$ (fifth and sixth columns for mirror and firehose respectively). The top row is for $a = 0.98$ while the bottom row is for $a = 0.5$. Grey regions indicate $\sigma \geq \sigma_{\rm cut}=10$. Since MAD simulations are highly magnetized with low plasma-$\beta$ in much of the volume, the mirror (c and i, $\eta>1$) and firehose (d and j, $\eta<1$) instabilities allow for a large temperature anisotropy in much of the volume. The mirror case also results in a higher electron temperature $\epsilon_\perp$, while the firehose case results in a lower and more uniform $\epsilon_\perp$. 

Future observations aim to probe not only the direct emission from the BH but also the lensed emission associated with the ``photon ring'' \citep{ngEHT2023}. The latter can be decomposed into a series of sub-rings labeled by the ray order $n$ -- the number of half-orbits a photon traveled to the observer, defined as $(\Delta \phi_{\rm ray} \mod \pi)$, where $\Delta \phi_{\rm ray}$ is the change in the angular coordinate $\phi_{\rm ray}$ along the ray in the plane of its orbit. To distinguish $n=0$ (direct) and $n=1$ (``photon ring'' of order $1$) in the ray tracing, we track $dz/ d \lambda$ along each ray, where $z$ and $\lambda$ are Cartesian Kerr Schild coordinate along the spin axis and coordinate along the ray respectively. The number of times that $d z/ d \lambda$ crosses zero for a particular ray defines the order of this ray $n$, allowing us to approximately distinguish $n=0$ and $n=1$.

\subsection{230 GHz images}{\label{sec:230EHT}}
\begin{figure*}
    \centering
    \includegraphics[width=0.98\textwidth]{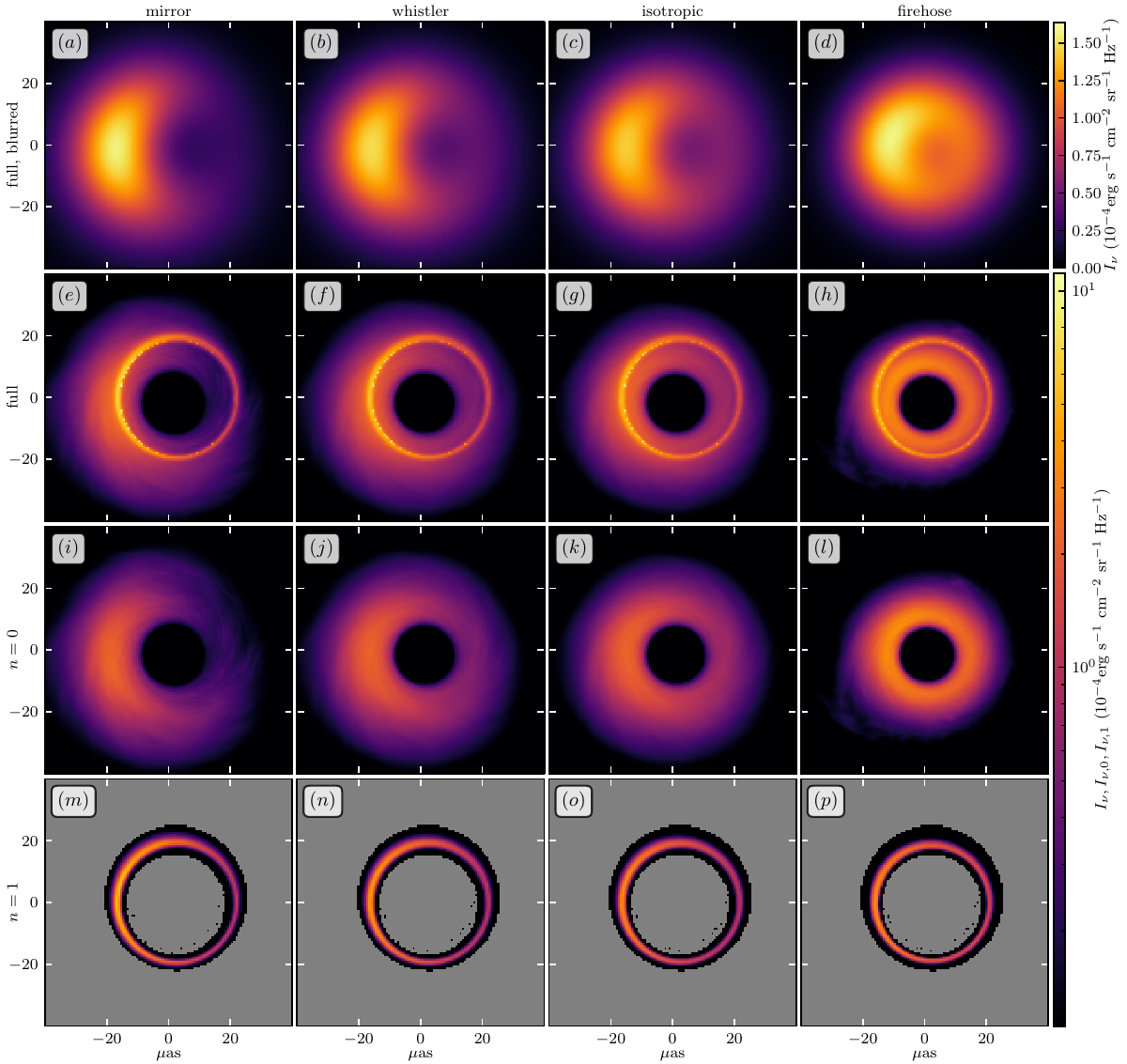}
    \caption{Synchrotron emission of accreting plasma raytraced from a MAD simulation for a BH with $a=0.98$ at $R_{\rm high}=10$ and inclination of $\theta=163^\circ$. Each column represents different plasma anisotropy: mirror instability threshold (column 1), whistler instability threshold (column 2), isotropic plasma distribution (column 3), and firehose instability threshold (column 4). The first row represents the full image blurred with $20 \mu$as FWHM Gaussian kernel on a linear scale (a-d), the second row shows a full unblurred image on a logarithmic scale, from which $I_{\nu,0}$ ($n=0$) and $I_{\nu,1}$ ($n=1$) are decoupled on the third (i-l) and forth (m-p) rows respectively.}
    \label{fig:theta163}
\end{figure*}
\begin{figure*}
    \centering
    \includegraphics[width=0.98\textwidth]{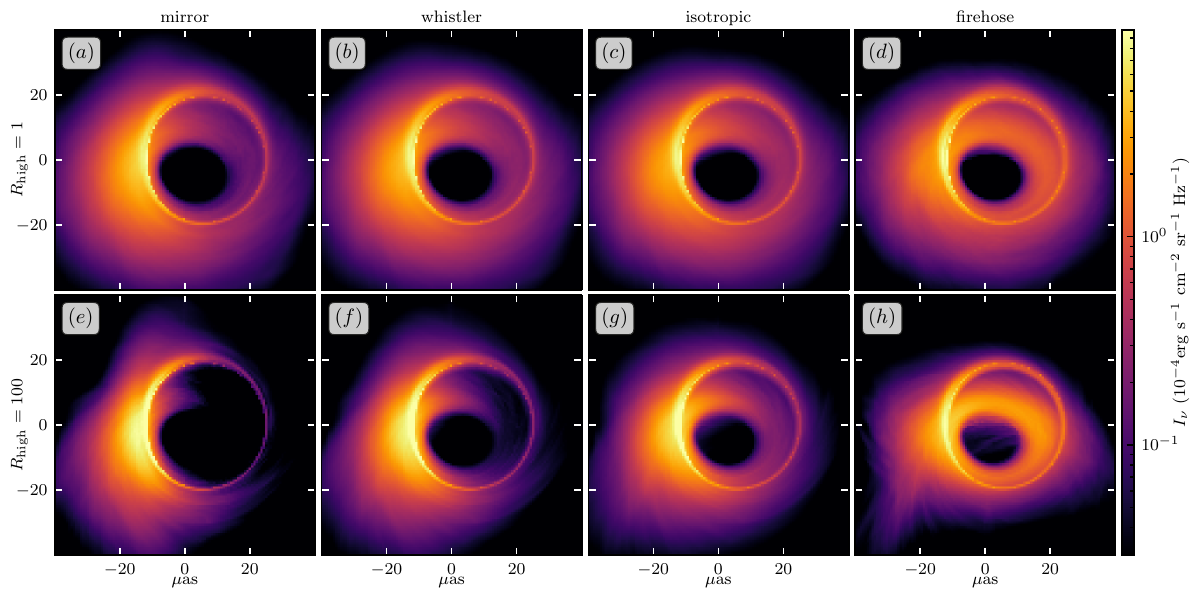}
    \caption{Synchrotron emission of accreting plasma raytraced from a MAD simulation at an inclination of $\theta=135^\circ$ for a BH with $a=0.98$. As in Fig.~\ref{fig:theta163}, each column represents different plasma anisotropy. The first and second rows represent $R_{\rm high}=1$ and $100$ models respectively.}
    \label{fig:theta135}
\end{figure*}
\begin{figure*}
    \centering
    \includegraphics[width=0.98\textwidth]{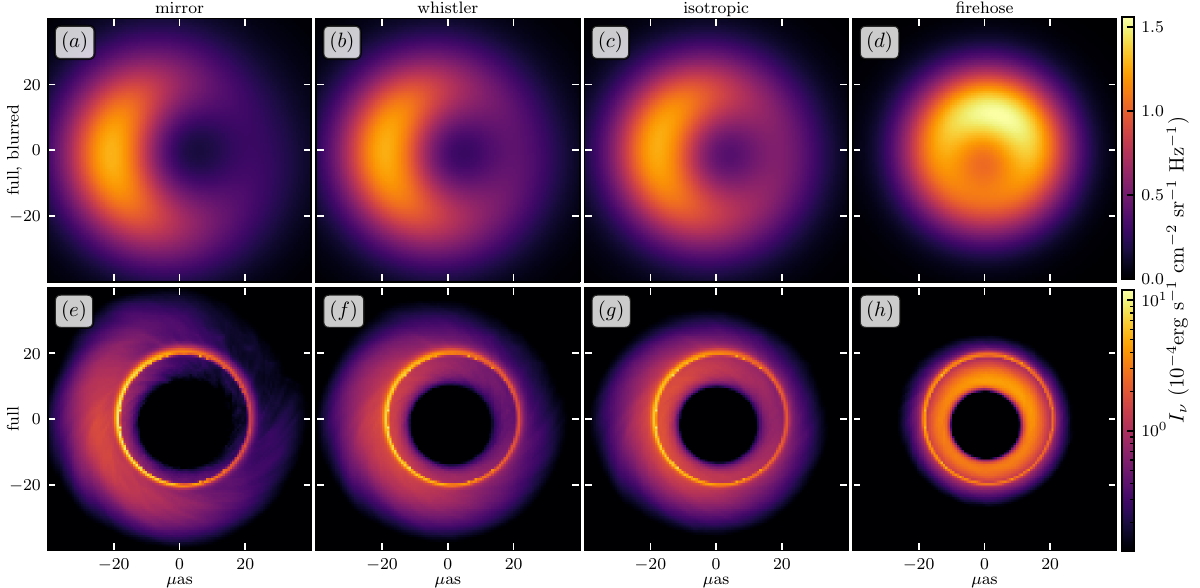}
    \caption{Similar to Fig.~\ref{fig:theta163} but for a BH with spin parameter of $a=0.5$. Inclination of $\theta=163^\circ$ and $R_{\rm high}=10$ are identical to Figure~\ref{fig:theta163}. The first row represents the full image blurred with $20 \mu$as FWHM Gaussian kernel on a linear scale (a-d), and the second row shows a full unblurred image on a logarithmic scale.}
    \label{fig:theta163_a05}
\end{figure*}
Total intensity images observed at $\theta=163^\circ$, expected for M87* \citep{M87angle2016A&A}, with $R_{\rm high}=10$ are shown in Figure \ref{fig:theta163} for $a=0.98$. The top row (a-d) shows the brightness blurred with $20 \mu as$ FWHM Gaussian kernel on a linear scale to match current EHT observations. Each column represents a different anisotropy model: mirror, whistler, isotropic, and firehose (from left to right, from largest to smallest $\eta$). The density normalization is roughly the same for each of these cases at fixed $R_{\rm high}$ and observing angle $\theta$, with the density in the firehose model being larger than in the isotropic case by a factor of a few. The three bottom rows in Figure \ref{fig:theta163} show unblurred full emission (second row), which is decomposed into the direct emission ($n=0$, third row) and the $n = 1$ photon ring (fourth row) on a logarithmic scale (as appropriate for future higher dynamic range measurements).

The azimuthal anisotropy in the images in Figure \ref{fig:theta163} is due to a combination of two effects: Doppler beaming and differences in the angle $\theta_B$ relative to the local magnetic field that photons are emitted at, in order to arrive at a given location in the observed image. Anisotropy in the electron distribution function can significantly change the synchrotron emission as a function of $\theta_B$, thus changing this second source of azimuthal image anisotropy. Figure \ref{fig:theta163} shows that, compared to the isotropic case (c), the mirror and whistler images [$\eta>1$, (a) and (b)] are more azimuthally asymmetric, while plasma at the firehose instability threshold [$\eta<1$, (d)] results in a more symmetric image. This is also noticeable in the unblurred case, as well as separately in $n=0$ and $n=1$ images. 

The dependence of the azimuthal image symmetry on plasma anisotropy is also more apparent with increasing viewing angle, i.e. as we look more ``edge-on'' instead of ``face-on''. Additionally, the effect of the anisotropy of the distribution function is more prominent for larger $R_{\rm high}$. This is because larger $R_{\rm high}$ suppresses the emission from high-$\beta$ regions (where distribution function anisotropies are constrained to be smaller) relative to low-$\beta$ regions (where distribution function anisotropies can be larger).  
The more azimuthally asymmetric image at higher inclination and higher $R_{\rm high}$ are demonstrated in Figure \ref{fig:theta135}, where we show the full intensity images at $R_{\rm high}=1$ (top) and $R_{\rm high}=100$ (bottom) at a higher inclination relative to the spin axis of $\theta=135^\circ$ for $a=0.98$.

{Table \ref{tab:average_eta} shows the image-averaged, emission-weighted ratio of the two components of anisotropic temperature, $\langle j_\nu T_\perp/T_\parallel \rangle / \langle j_\nu \rangle$, for $a=0.98$, $\theta=163^\circ$, $R_{\rm high}=1$, $10$, and $100$. As $R_{\rm high}$ increases, the anisotropic temperature ratio approaches our maximum allowed values of $10$ ad $0.1$ for mirror and firehose models respectively.  The significant changes in image morphology found here thus require large temperature anisotropy in the emitting plasma.} 

\begin{table}[]
    \centering
    \begin{tabular}{c c c c}
          $R_{\rm high} $& mirror & whistler & firehose\\
         \hline
         \hline
         1   & 3.13 & 1.47 & 0.52\\ 
         10  & 5.3 & 2.0 & 0.1100 \\ 
         100 & 8.9 & 2.9 &  0.102 \\ 
    \end{tabular}
    \caption{Values of image-averaged emission-weighted temperature anisotropy $T_\perp/T_\parallel$ [pixel-averaged $\langle j_\nu T_\perp/T_\parallel \rangle / \langle j_\nu \rangle$] for our three temperature anisotropy cases at a = 0.98, an inclination of $\theta=163 \deg$, and three $R_{\rm high}$ values.}
    \label{tab:average_eta}
\end{table}

To better understand the interplay between Doppler-induced asymmetry and magnetic field viewing angle-induced asymmetry, we also consider the case of a moderately spinning BH, $a=0.5$, shown in Figure \ref{fig:theta163_a05}, where the Doppler effect is smaller than for $a=0.98$ studied above. This figure is organized identically to Fig.~\ref{fig:theta163} and the viewing angle relative to the spin axis and the choice of $R_{\rm high}=10$ are the same. We find that the asymmetry of the image due to the plasma temperature anisotropy is still pronounced, similar to the case of a highly spinning BH. As in the $a=0.98$ case, mirror and whistler anisotropies make the image more asymmetric, while temperature anisotropy near the firehose boundary results in a more symmetric image.

Our calculations show that the anisotropic temperature distribution of plasma sitting at the firehose and mirror thresholds leads to a more azimuthally symmetric or asymmetric synchrotron image, respectively. At first glance, it is not entirely obvious why the firehose sense of anisotropy (rather than the mirror sense of anisotropy) should be associated with a more symmetric image. Our interpretation of this is that if the rotation rate of the magnetic field lines is small relative to the rotation rate of the plasma, then in ideal MHD models, the plasma velocity is approximately parallel to the magnetic field direction (see, e.g., eq. E148b of \citealt{Chael2023} for a relativistic version of this expression). For nearly (but not exactly) face-on viewing angles, the Doppler effect and the effect of changing viewing angle relative to the magnetic field are ``in phase'':  the brightening and dimming produced by the two effects peak in roughly the same places in the image plane (this follows, e.g., from the analytic model in \citealt{Narayan2021}). The firehose instability sense of anisotropy counteracts this by making the emission a significantly weaker function of angle relative to the magnetic field (Fig. \ref{fig:jnu}) thus making the overall emission more isotropic. 

Another key difference between images with different electron temperature anisotropy is the image diameter; this is noticeable at both spin values in Figures~\ref{fig:theta163} and \ref{fig:theta163_a05}: the size of the bright region in the image increases as $\eta$ increases. Additionally, $a=0.5$ shows variations in the size of the BH shadow between different models in Fig.~\ref{fig:theta163_a05}. Both of these effects, as well as the asymmetry of the images, are quantified below.

\begin{figure}
    \centering
    \includegraphics[width=0.98\columnwidth]{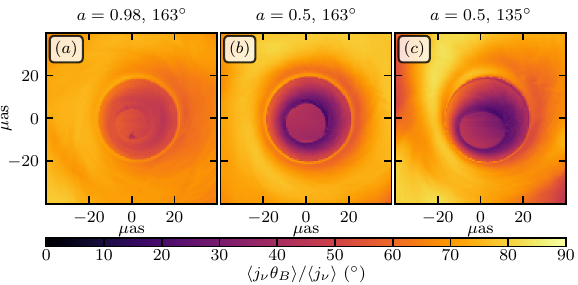}
    \caption{Average angle with the magnetic field along the ray, measured by emission-weighted sine of $\theta_B$ for different spins and viewing angles. Lower spin decreases the average angle of the emitted photons relative to the magnetic field. This in turn enhances the effects of plasma anisotropy on the observed image (Figures~\ref{fig:theta163}-\ref{fig:theta163_a05}).}
    \label{fig:angle}
\end{figure}

Figure~\ref{fig:angle} shows the emissivity-weighted angle between the magnetic field and photon direction along the ray $\langle j_\nu \theta_B \rangle / j_\nu$. This angle is larger for $a=0.98$ (a) than for $a=0.5$ (b,c) in the inner region of the image. Physically for roughly face-on viewing angles, the magnetic field in the accretion flow onto a BH with a smaller spin has a more vertical field than onto a highly spinning BH (where the field is wrapped up to be more azimuthal). This leads to the average angle between the propagation direction and the local magnetic field decreasing for $a=0.5$ relative to $a=0.98$. A less face-on viewing angle produces a similar effect (c). Figure~\ref{fig:angle} shows results for the isotropic emission model but can be used to gain insight into why the central ``shadow'' is noticeably different in the firehose and mirror cases in Figures~\ref{fig:theta163} and \ref{fig:theta163_a05}. In particular, the lower average angle between the photon and magnetic field in Figure~\ref{fig:angle} at lower spin and observer viewing angle implies (via Figure~\ref{fig:jnu}) that in the mirror (firehose) case the emission in the shadow should be suppressed (enhanced). This is exactly what is seen in the images. Plasma anisotropy could thus have an effect on observational efforts to infer physical properties of the black hole using the ``inner shadow'' \citep{Chael2021}.

We now quantify the effects of changing image size and asymmetry for different plasma anisotropy models. Following \citep{EHTm87PaperIV}, we measure the image diameter $d$ as twice the distance from the center of the image to the peak $I_\nu$ averaged over all directions and $w$ is the Full Width Half Maximum (FWHM) of $I_\nu$ averaged over all directions. We can then infer $r_{\rm in}=(d-w)/2$ and $r_{\rm out}=(d+w)/2$ -- inner and outer radius of the image. The asymmetry parameter $A$ of the image, defined in image plane coordinates $r_{\rm im}-\phi_{\rm im}$, is
\begin{equation}
    A = \left\langle \frac{\int_0^{2\pi} I(\phi_{\rm im}) e^{i \phi_{\rm im}} d \phi_{\rm im} }{\int_0^{2\pi }I(\phi_{\rm im}) d \phi_{\rm im}} \right\rangle_{r_{\rm im} \in [r_{\rm in}, r_{\rm out}]},
\end{equation}
where $I(\phi_{\rm im})$ is the brightness profile across image coordinate $\phi_{\rm im}$ at a fixed radial coordinate $r_{\rm im}$. A fully symmetric image has $A=0$, while an antisymmetric image has $A=1$. 

\begin{figure}
    \centering
    \includegraphics[width=\columnwidth]{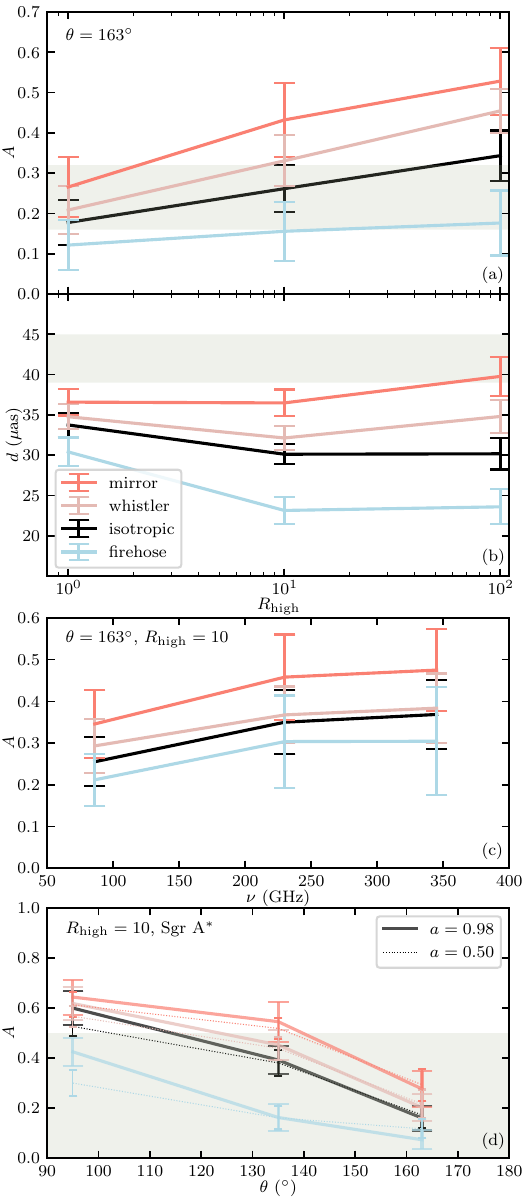}
    \caption{Asymmetry $A$ (a) and diameter $d$ (b) as functions of $R_{\rm high}$ for $a=0.98$ and M87* observing angle, $\theta=163^\circ$, for images at $230$ GHz blurred with $20\mu$as FWHM Gaussian kernel. (c): Asymmetry of unblurred images at $\theta=163^\circ$ and $R_{\rm high}=10$ as a function of observing frequency, $a=0.98$. (d): Asymmetry at $R_{\rm high}=10$ as a function of observing angle $\theta$ for $a=0.98$ (solid lines) and $a=0.5$ (thin dotted lines) for Sgr A*. The green regions highlight EHT constraints for M87* (a,b) and Sgr A* (d). In each panel, the color of the lines represents 4 limiting cases: mirror instability, whistler instability, isotropic plasma distribution, and firehose instability.}
    \label{fig:asymmetry}
\end{figure}
The asymmetry $A$ and diameter $d$ measured from raytraced images are shown in Figure~\ref{fig:asymmetry} with different models represented by different colors, identical across all panels; $230$ GHz results and the variation with frequency are shown in panels (a,b,d) and (c) respectively. Top panels show $A$ (a) and $d$ (b) for an M87* observing angle $\theta=163^\circ$ as functions of $R_{\rm high}$. Panel (d) shows $A$ for $R_{\rm high}=10$ as a function of observing angle $\theta$ for $a=0.98$ (solid lines) and $a=0.5$ (thin dotted lines) for Sgr A*. Shaded regions indicate the allowed range as inferred from observations for M87* (a-b) \citep{EHTm87PaperIV, EHTm87PaperVI} and Sgr A* (d) \citep{EHTsgrPaperIV}. Panel (c) shows $A$ measured from unblurred images as a function of frequency for an M87* viewing angle. 

As expected, the difference in anisotropy $A$ between the models becomes larger with increasing $R_{\rm high}$ (a) since larger $R_{\rm high}$ suppresses the emission from high-$\beta$ regions relative to low-$\beta$ regions, where the plasma can develop significant anisotropy (see also Table~\ref{tab:average_eta}). The firehose case ($\eta<1$) always shows smaller $A$, consistent with the images in Figures~\ref{fig:theta163}-\ref{fig:theta163_a05}, while models with $\eta>1$ show higher $A$ compared to the isotropic case. The firehose models typically have anisotropy $A$ up to $\lesssim 3$ smaller than the mirror case, with the exact value depending on $R_{\rm high}$ and viewing angle. As explained above, this is because plasma at the firehose limit ($\eta<1$) emits more isotropically (over a wide range of angles) with respect to the magnetic field direction, relative to the mirror case which emits mostly at $\theta_B = 90^\circ$.  This leads to less anisotropy in the image overall. In M87* the viewing angle is constrained to be $\theta\approx163^\circ$, while $A\approx0.16-0.32$; thus, from Figure \ref{fig:asymmetry}a, a better fit to the observed $A$ is obtained for $\eta<1$ at larger $R_{\rm high}$ or $\eta>1$ with smaller $R_{\rm high}$. 

We also show the diameter of the image in Figure~\ref{fig:asymmetry}b, calculated for the same images as shown in panel (a), i.e., $a=0.98$, $\theta=163^\circ$, and $R_{\rm high}=10$; the diameter is generally larger for models with larger $\eta$. The shaded region indicates M87* constraint of $d= (42 \pm 3) \mu$as \citep{EHTm87PaperVI}. As was shown in Figure~\ref{fig:epsilon}, the temperature $\epsilon_\perp$ for plasma at the firehose limit (f,l) is smaller and varies less with radius than at the mirror limit (e,k) for both spin parameters. This lower temperature leads to emission more concentrated near the black hole and thus a smaller image diameter.

Figure~\ref{fig:asymmetry}d shows that the image becomes more asymmetric (d) as we look more ``edge-on'' instead of ``face-on''.  Both spin values of $0.98$ (solid lines) and $0.5$ (thin dotted lines) show similar behavior. The images used for panel (d) are produced for Sgr A* with $M = 4.3 \times 10^6 M_\odot$, and the density is normalized such that the total flux matches EHT observations, i.e., $F_\nu = 2.4$Jy at $230$ GHz and distance of $d = 8178$pc. A quantitatively similar trend, however, is also present for our M87* models. We also show the EHT constraints on $A$ for Sgr A* by the shaded region in (d) [$A \approx 0-0.5$]. As before, plasma at the firehose anisotropy limit leads to a more symmetric image, compared to mirror and whistler limits, at any observing angle. Note the quite isotropic image (small A) at the firehose limit even at $\theta=135^\circ$, especially for the lower spin case $a=0.5$. This effect can significantly change the constraint on our viewing angle relative to Sgr A* suggested by the EHT data, allowing for larger observing angles than for an isotropic plasma.

\begin{figure}
    \centering
    \includegraphics[width=\columnwidth]{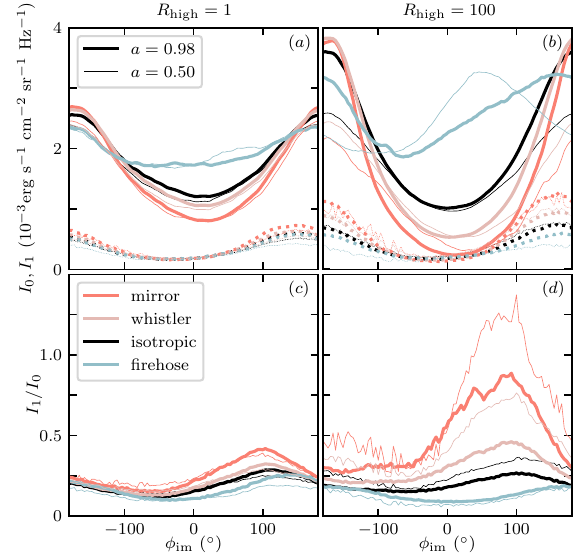}
    \caption{Angular profiles of $n=0$ and $n=1$ brightness ($I_0$ [solid] and $I_1$ [dotted] as functions of $\phi_{\rm im}$ in the image plane, top row) and their ratio (bottom row) at observing angle of $\theta=163^\circ$ at spin of $0.98$ (thick lines) and $0.5$ (thin lines). The first and second columns represent $R_{\rm high}=1$ and $100$ respectively. The color of the lines, as in Fig.~\ref{fig:asymmetry}, represents 4 limiting cases: mirror, whistler, isotropic, and firehose.}
    \label{fig:Iphi}
\end{figure}
We will now quantify the imprint of the anisotropy of the plasma distribution function on the direct emission $I_0$ and the $n = 1$ photon ring $I_1$ separately. As seen in Fig.~\ref{fig:theta163}~(i-p), both $n=0$ and $n=1$ emission have their azimuthal asymmetry modified with varying $\eta$ in a way that is similar to the full blurred image. Both are more symmetric at the firehose limit with $\eta<1$ and more asymmetric at the mirror and whistler limits with $\eta>1$, compared to the isotropic plasma distribution case. To distinguish the imprint of the plasma anisotropy on the two components, we show angular profiles of $n=0$ and $n=1$ emission ($I_0$ and $I_1$ as functions of the polar angle in the image plane $\phi_{\rm im}$, top row, a and b) and their ratio ($I_1/I_0$, bottom row, c and d). The polar angle is plotted such that the dimmest region of the image, $\phi_{\rm im}\sim0$, is in the center of the profile. This is for an observing angle of $\theta=163^\circ$ for both of our spin values of $0.98$ and $0.5$ (thick and thin lines, respectively); $R_{\rm high}=1$ and $100$ are shown in the left and right columns respectively.

As expected, the $R_{\rm high}=100$ case shows a stronger dependence of $I_0$ and $I_1$ on plasma anisotropy than $R_{\rm high}=1$ due to the higher anisotropy in the low-$\beta_{\rm th}$ regions.  The quantitative dependence of the $n = 0$ and $n = 1$ intensities on plasma anisotropy differ because the $n = 0$ and $n = 1$ photons at the same place in the image plane are emitted at different directions relative to the local magnetic field. The largest difference between $I_0$ and $I_1$ is reached in the case of smaller electron temperatures at the mirror limit. In principle, measurements of the azimuthal intensity profiles at $n = 0$ and $n = 1$ could thus be used to constrain plasma anisotropy though it is unclear if this is feasible in practice given uncertainties in black hole spin, the electron temperature, degree of Doppler beaming, etc. 

In addition to calculating the synchrotron emission and absorption produced by an anisotropic distribution function, we have also calculated how the emitted linear and circular polarization depends on plasma anisotropy. Because we do not consider the impact of plasma anisotropy on Faraday rotation and conversion in this paper, we defer a detailed discussion of the polarization due to plasma anisotropy to future work. We can, however, quantify the change in intrinsic linear and circular polarization, i.e. neglecting the effects of Faraday rotation and conversion. We find that the image-averaged linear polarisation fraction can change by up to roughly $+10$\% or $-10$\% for the mirror and firehose limits respectively, compared to the isotropic case. Circular polarization exhibits the same trend, but the mirror case can be $5$ times more circularly polarized compared to the isotropic case, at $R_{\rm high}=100$. We also note that because models with plasma at the firehose anisotropy have smaller $\epsilon_\perp$, a higher density is required to match the observed EHT flux. This leads to an increase in pixel-averaged optical depth, e.g.: $1.1 \times 10^{-3}$, $1.2 \times 10^{-3}$, $1.3 \times 10^{-3}$, and $3.7 \times 10^{-3}$ for the mirror, whistler, isotropic, and firehose cases respectively at an inclination of $163^\circ$ and $R_{\rm high}=10$. Thus, $\tau$ is by a factor of $3-4$ larger in the firehose case, compared to other cases, which might also lead to a higher Faraday depolarization.

\subsection{Multi-wavelength observations}\label{sec:freq}
\begin{figure}
    \centering
    \includegraphics[width=0.98\columnwidth]{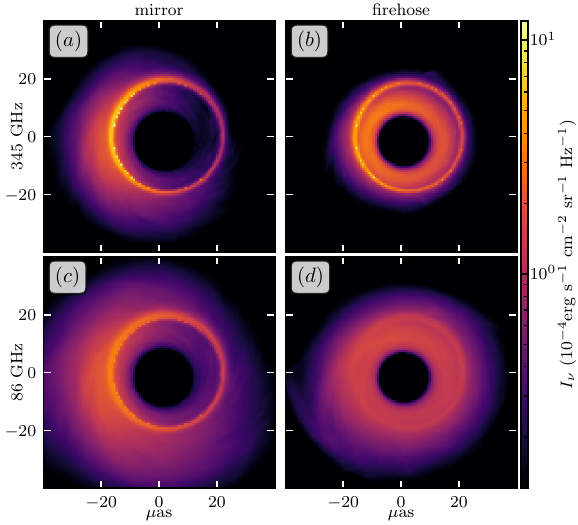}
    \caption{Synchrotron emission of accreting plasma raytraced from a MAD simulation with $a=0.98$ at an inclination of $\theta=163^\circ$ and $R_{\rm high}=10$ at frequencies of $345$ GHz (a-b) and $86$ GHz (c-d). The first and second columns represent two limiting cases (mirror and firehose respectively).}
    \label{fig:freq}
\end{figure}
\begin{figure*}
    \centering
    \includegraphics[width=\textwidth]{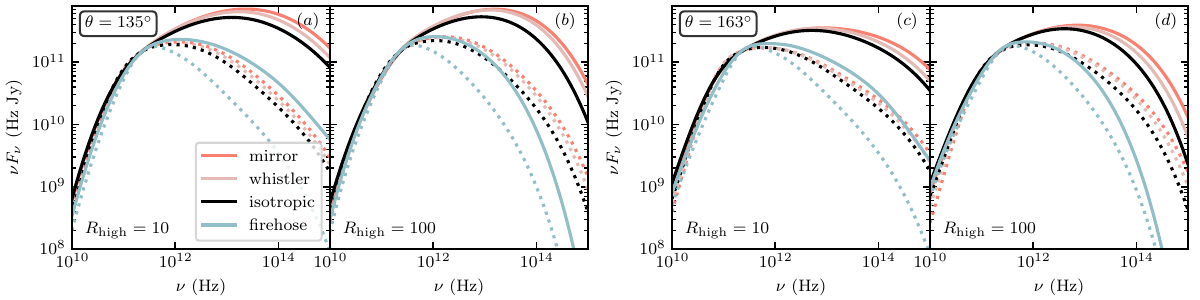}
    \caption{Synchrotron emission spectra for BHs with $a=0.98$ (solid lines) and $a=0.5$ (thin dotted lines) viewed at inclinations of $\theta=135^\circ$ (a,b) and $163^\circ$ (c,d) at $R_{\rm high}=10$ (a,c) and $100$ (b,d). As in Fig.~\ref{fig:asymmetry}, different models are represented by different colors. }
    \label{fig:spectra}
\end{figure*}

Future mm interferometric observations will include 2 more frequencies, $345$ GHz and $86$ GHz \citep{ngEHT2023}, with the latter (former) expected to be more (less) optically thick \citep{Chael2023ApJfreq}. We thus explore the impact of an anisotropic plasma distribution function on observable images and spectra at these frequencies. In Figure~\ref{fig:freq} we show intensity images for a BH with $a=0.98$ at $345$ GHz on top (a-b) and $86$ GHz on the bottom (c-d), with the parameters being identical to Figure~\ref{fig:theta163} -- $\theta=163^\circ$ and $R_{\rm high}=10$. The mirror and firehose models are shown in the first (a,c) and second (b,d) columns respectively. The respective images at $230$ GHz are shown in Figure~\ref{fig:theta163} for mirror (e) and firehose (h) cases. 

The differences between the mirror and firehose in Figure~\ref{fig:freq} at $345$ GHz are similar to the differences at $230$ GHz in Figure~\ref{fig:theta163}: the mirror case is more azimuthally asymmetric than the firehose case. Images at $345$ GHz (a-b) are particularly similar to their $230$ GHz counterparts because the emission is predominantly optically thin in both cases. At lower frequency (c-d), however, the higher synchrotron optical depth somewhat suppresses the differences between the mirror and firehose limits and overall makes the emission more azimuthally symmetric. Figure~\ref{fig:asymmetry}c quantifies the asymmetry $A$ as a function of frequency for the 4 different distribution function models -- the difference in the asymmetry between the different distribution function models persists at all frequencies though the overall asymmetry is largest at high frequencies. In the firehose model at $86$ GHz, Figure~\ref{fig:freq}d also shows that the photon ring emission is much less evident. This is because the firehose model has a lower temperature and higher density (at fixed $230$ GHz flux) than the other plasma anisotropy models, and so the emission is optically thick at $86$ GHz. {The same trend, i.e. optically thin emission at high frequencies ($345$ GHz and $230$ GHz) and optically thick emission at $86$ GHz in the firehose case, persists at a lower spin parameter of $a=0.5$ (not shown here).}

We also calculate the synchrotron emission spectra from $10^{10}$ to $10^{15}$ Hz, shown in Figure \ref{fig:spectra} at $135^\circ$ (left, a-b) and $163^\circ$ (right, c-d) for a spin of $0.98$ (solid lines) and $0.5$ (dotted lines) at two $R_{\rm high}$ values of $10$ (a,c) and $100$ (b,d) (shown on the left and right side panels for each angle respectively). The different spectra for different black hole spins are due to the higher temperatures found in more rapidly spinning GRMHD simulations \citep{Moscibrodzka2009ApJ}. The color of the lines is organized as in previous plots with different colors representing different plasma anisotropy. The firehose case shows a significantly different spectrum for both $a=0.98$ and $a=0.5$. The change is minor at low frequencies, with firehose being slightly fainter than the other models. The peak of the spectrum, however, can significantly shift to lower frequencies, steepening the spectral slope just below the peak. At higher frequencies, the emission in the firehose model is substantially fainter and the spectral slope is steeper, compared to other cases. The qualitative results do not depend on the value of $R_{\rm high}$. Our physical interpretation of this is that at fixed GMRHD temperature, the firehose model (with $T_{\parallel,e} > T_{\perp,e}$) has a lower value of $T_{\perp,e}$. This suppresses the peak frequency of the synchrotron emission as given by Equation~\ref{eq:nu_peak} leading to a more rapid decline in emission at high frequency.

\section{Summary and Conclusions}\label{sec:sum}
Magnetized collisionless plasmas are prone to developing anisotropies in their distribution function with respect to the magnetic field direction: the distribution function is isotropic in the plane perpendicular to the magnetic field because of rapid cyclotron motion (``gyrotropic''), but can be very different along and perpendicular to the local magnetic field. In this work we have calculated the synchrotron radiation from distribution functions with anisotropy of this form. We are motivated by the application to low accretion rate black holes such as those found in Sgr A* and M87* but we anticipate that the synchrotron radiation calculations presented here will have broader applicability.

First, we have derived and provided fits for synchrotron emissivities and absorption coefficients for relativistic thermal electrons with an anisotropic distribution function in (Eq.~\ref{eq:fits}).  The distribution function we choose (Eq.~\ref{eq:df}) is a natural relativistic generalization of a non-relativistic bi-Maxwellian and allows for arbitrary temperature anisotropies relative to the local magnetic field $T_\perp/T_\parallel$ via a parameter $\eta$. The derived fits we present are accurate to $\sim 10 \%$ or better compared to numerical solutions using the publicly available synchrotron code {\tt symphony} \citep{Pandya2016ApJ} in the parameter range of interest (high frequency and high temperature); the main source of error is the inaccuracy of the fits for synchrotron emission and absorption for an isotropic thermal plasma, which our fits are scaled to.

The change in synchrotron emission as the plasma transitions from an isotropic to anisotropic distribution function at a fixed perpendicular temperature $T_\perp$ can be understood as a renormalization of the number of particles that emit toward the observer. The reason is that synchrotron emission emitted at an angle $\theta_B$ relative to the local magnetic field is produced primarily by particles whose vector momenta are in the same direction as $\theta_B$, or, equivalently, the pitch angle of the emitting particles is $\xi \approx \theta_B$. The emission thus depends on the distribution function at pitch angle $\xi \approx \theta_B$. For a plasma with an isotropic distribution function the temperature is independent of $\xi$ but temperature anisotropy in the distribution function implies that the temperature is now effectively a function of pitch angle $\xi$ and thus viewing angle $\theta_B$ (Eq.~\ref{eq:epsperpstar}). For an isotropic plasma, synchrotron emission is peaked near $\theta_B \sim 90 \deg$, i.e., orthogonal to the magnetic field. This trend is {\em enhanced} for $T_\perp > T_\parallel$ (anisotropy parameter $\eta > 1$) while for $T_\parallel > T_\perp$ ($\eta<1$) the emission can peak at significantly smaller observing angles, depending on the exact value of $\eta$ (Fig.~\ref{fig:jnu}). The case of $\eta<1$ also shows more uniform emission across observing angles than does $\eta>1$.

In addition to calculating the total emitted synchrotron radiation as a given frequency, we have also calculated the emitted linear and circular polarization fractions as a function of plasma anisotropy. We find that the intrinsic linear polarization degree depends only weakly on the plasma anisotropy $\eta$. On the other hand, circular polarization, which is very weak in synchrotron emission from relativistic isotropic plasmas, increases significantly for $T_\parallel < T_\perp$ at a fixed $T_\perp$ ($\eta>1$). In addition, since most of the emission comes from large (small) angles relative to the magnetic field for $\eta>1$ ($\eta<1$), the respective angle-averaged circular polarization degree is higher (smaller).

We have employed the newly developed fits for synchrotron emission and absorption by anisotropic electrons in a GR radiative transfer code {\tt blacklight}, capable of propagating synchrotron radiation in curvilinear space-time. To assess how anisotropy of the accreting plasma affects mm-wavelength observations of Sgr A* and M87*, we ray-trace GRMHD MAD simulations -- the accretion model most favored observationally \citep{EHT2021ApJmagneticField}. {Other accretion models, such as Standard and Normal Evolution (SANE) models, are also possible. In such models the plasma-$\beta$ is considerably higher. This suggests that the effect of plasma anisotropies is relatively smaller in SANE models compared to MAD models, but more detailed work in the future is required to assess this quantitatively.} Since the ideal MHD approach describes a collisional isotropic fluid, the main source of uncertainty in this work is the temperature and temperature anisotropy of the synchrotron-emitting electrons. In particular, the ion-to-electron temperature ratio, which we approximate by the widely-used $R_{\rm high}-R_{\rm low}$ model and the electron's anisotropy $\eta$, are the main free parameters in our study. Since $\eta$ is a prescribed quantity, absent in our ideal GRMHD simulations, the conclusions of this work should be thought of as qualitative rather than quantitative.

The temperature anisotropy in a collisionless plasma cannot grow without bound because small-scale instabilities set in and limit the magnitude of the temperature anisotropy. We thus examine the effect of an anisotropic synchrotron emitting plasma on observed emission by considering 3 limiting cases, defined by the anisotropy thresholds of three anisotropy-driven instabilities: the mirror and whistler instabilities ($\eta>1$) and the firehose instability ($\eta < 1$). We present relativistic derivations of these thresholds in Appendix~\ref{ap:instability}. In particular, we derive a fully kinetic mirror instability threshold in the case of anisotropic relativistic electrons with anisotropy parameter $\eta$ (and anisotropic non-relativistic ions with a different anisotropy parameter $\eta_i$). The temperature anisotropy allowed by kinetic-scale instabilities is larger for stronger magnetic fields, i.e., smaller $\beta$ (the ratio of thermal to magnetic energy density). The effects of temperature anisotropy on observed synchrotron emission are thus likely to be the largest when the emission is dominated by regions with $\beta \lesssim 1$, as is often the case in magnetically-arrested disk models favored on theoretical and observational grounds.


We find that anisotropy in the accreting plasma can significantly modify the observed synchrotron emission in horizon-scale images, including the azimuthal asymmetry in the image plane and size of the image. This is primarily due to the following two effects. The first effect is that the emission and absorption for different distribution anisotropies are concentrated at different observing angles with the magnetic field, with $\eta<1$ emitting more uniformly across all angles as $\eta$ decreases, and $\eta > 1$ emission/absorption being more concentrated near $\theta_B \sim 90 \deg$ (Fig.~\ref{fig:jnu}). This can significantly modify the azimuthal asymmetry in the image plane because different parts of the image contain radiation that was initially emitted at different angles relative to the local magnetic field. The second key effect is that the local perpendicular temperature $T_\perp$ of the electrons changes with an assumed anisotropy $\eta$ at a given total fluid temperature $T$ given by the GRMHD solution (Fig.~\ref{fig:epsilon}). Models with $\eta > 1$ ($\eta<1$) have a larger (smaller) $T_\perp$, compared to the isotropic case. Higher (lower) temperatures produce larger (smaller) 230 GHz images because the emission at 230 GHz occurs over a larger (smaller) range of radii (Fig.~\ref{fig:theta163}). Higher temperatures also lead to a smaller (higher) density of the accreting plasma at a fixed $230$ GHz flux and thus more optically thin (thick) emission; this is especially pronounced for $\eta>1$, i.e., the firehose regime, in which the image-averaged optical depth can increase by a factor of $3-4$.

More specifically we find that emission from plasma with $\eta<1$ ($\eta>1$) produces a more azimuthally symmetric (asymmetric) image, up to a factor of $3$ difference in the asymmetry parameter $A$. This result is of particular interest in application to Sgr A*, where the observed EHT azimuthal asymmetry is surprisingly modest given expectations for a random viewing angle. This appears to suggest we are observing Sgr A* closer to face-on than not, which is a priori surprising. Models with $\eta < 1$ have significantly less variation in the synchrotron emissivity with photon direction relative to the magnetic field. This produces a more azimuthally symmetric image, alleviating the restrictive constraints on viewing angle (Fig.~\ref{fig:asymmetry}d).

Anisotropy in the plasma distribution function also changes the image diameter and the size of the central flux depression (or the observed ``BH shadow''). The smaller perpendicular temperature $T_\perp$ in $\eta<1$ firehose model results in a reduced image diameter (Fig.~\ref{fig:asymmetry}b). At lower BH spins, the viewing angle relative to the magnetic field is also smaller in the near-horizon region. This suppresses (enhances) the emission in the image center interior to the true photon ring (i.e., the critical curve). The BH ``shadow'' therefore appears to be larger in low spin models with $\eta>1$ (Fig.~\ref{fig:theta163_a05}). \citet{Chael2021} showed that the size and shape of the ``inner shadow'' depend on BH spin and our viewing angle relative to the BH spin, potentially providing a route to measuring these quantities. Our results show that anisotropy in the distribution function in this region close to the event horizon may be important to consider as well.

In this paper we have not calculated the Faraday conversion coefficients for an anisotropic plasma. We defer this to future work. We have, however, calculated the emitted linear and circular polarization fractions and how they depend on plasma anisotropy. We find that the imaged-averaged emitted linear polarization fraction can increase (decrease) by up to $10\%$ in the mirror and whistler (firehose) cases. The emitted circular polarization fraction shows a similar trend, although the magnitude of the effect is much larger, with the $T_\perp > T_\parallel$ regime showing an emitted circular polarization in the mm that is up to $5$ times larger than in an isotropic plasma. 

{ The high frequency synchrotron emission is particularly sensitive to plasma anisotropy.  As a result, interpreting and modeling  GRAVITY observations of Sgr A* may require incorporating the effects of plasma anisotropy; this emission is also likely non-thermal, however, so an extension of our results to non-thermal distribution functions would be valuable.}

We have also assessed how the anisotropy of the plasma affects future multi-frequency and $n = 1$ photon ring observations. We find that the effect of the plasma distribution function on the azimuthal image asymmetry persists throughout the frequencies of interest to future ngEHT observations, i.e. $86$ GHz and $345$ GHz, though the effect is more pronounced at higher frequencies (Fig.~\ref{fig:asymmetry}c).  
We also find that the $n = 1$ photon ring emission is even more azimuthally asymmetric (symmetric) for $\eta>1$ ($\eta<1$) than the direct $n = 0$ emission, leading to an increased (decreased) ratio of photon ring to direct emission brightness -- up to a factor of $6$ in intensity ratio relative to the isotropic distribution function case for the parameter range we considered. Anisotropy in the distribution function has a particularly large effect on the ratio of the $n = 1$ to $n = 0$ emission because plasma anisotropy directly changes the emissivity as a function of viewing angle relative to the magnetic field, and the $n = 0$ and $n = 1$ images contain emission emitted at different angles relative to the local magnetic field.

The largest limitation of the present study as applied to modeling Sgr A*, M87* and related sources is that the true electron temperature anisotropy in the near-horizon environment is poorly constrained. In this work we have attempted to bracket the magnitude of the effect that temperature anisotropy can produce on near-horizon synchrotron radiation by considering the extreme limit in which all of the plasma is at the temperature-anisotropy associated with the instability thresholds for the mirror, whistler, or firehose instabilities. The image-averaged emission-weighted electron temperature anisotropies in these models are given in Table \ref{tab:average_eta} and range from $\sim 0.1-9$. Real systems likely do not follow just one of the limiting anisotropy models considered here since different temperature anisotropy can co-exist in different parts of the accretion flow. {In magnetically dominated jet regions, the plasma is in principle capable of developing large anisotropy in its distribution function.   This could occur due to differential parallel and perpendicular heating and/or as a result of outflow-driven expansion of the jet (as in the solar wind).  Consequently, it would be interesting to apply the methods developed here -- likely extended to non-thermal distribution functions -- to model and interpret the emission from spatially extended jets (e.g., \citealt{jet2023Nature}).}    

Fortunately, there is a clear path forward for improving our understanding of the role of temperature anisotropy in the radiation from accretion flows and jets.  Global ``extended'' MHD models that evolve the pressure anisotropy as a dynamical variable can predict $T_\perp/T_\parallel$ as a function of time and space \citep{Foucart2017MNRAS}, removing the need to specify the temperature anisotropy in post-processing as we have done here; such models will, however, needed to be extended to consider both electron and proton temperature anisotropies.  Global GRPIC simulations can go one step further and predict the full distribution function in the accretion flow and outflow, including temperature anisotropy, and deviations from a Maxwellian, etc. \citep{GRPIC2023PhRvL}. One aspect that is important to account for in future modeling is that in plasmas with $T_p > T_e$, the mirror and fluid firehose instabilities are most sensitive to the proton temperature anisotropy (see Appendix \ref{ap:instability}). As a result it is plausible that the electron anisotropy is set primarily by resonant instabilities such as the whistler and resonant firehose instabilities.

\begin{acknowledgments}
We thank Alex Lupsasca for useful conversations and for sharing his analytical calculations of source-plane emission angle for different photon sub-rings. We also thank Chris White for useful conversations and for sharing the details of his GRMHD MAD simulations, and Charles Gammie and Michael Johnson for useful conversations. This research was facilitated by the Multimessenger Plasma Physics Center (MPPC), NSF grant PHY-2206610. A.P.~ acknowledges support by NASA ATP grant 80NSSC22K1054. This work was also supported in part by a Simons Investigator Grant from the Simons Foundation (EQ), and was completed during EQs stay at the Aspen Center for Physics, which is supported by National Science Foundation grant PHY-2210452.
\end{acknowledgments}

\appendix

\section{Comparison of analytical synchrotron expressions with numerical results}\label{ap:comparison}

In this Appendix we analytically calculate the synchrotron emission and absorption coefficients for our assumed gyrotropic distribution function and compare the resulting analytic expressions to full numerical evaluations of Equations~\ref{eq:S_equations}. The analytic calculations are carried out in the limit of high Lorentz factors for the emitting electrons, the same regime in which analytic progress can be made for an isotropic distribution function [see, e.g., \citep{GinzburgSyrovatskii1965, Melrose1971}].

\subsection{Derivation of Analytical Fits for Total Intensity, Linear Polarization, and Circular Polarization}
Under the assumption of high Lorentz factor $\gamma$ (or energy $E$) for the emitting electrons, the emission is predominantly concentrated in a narrow cone around the pitch angle $\mu \simeq \cos(\theta_B)$ where $\theta_B$ is the viewing angle with respect to the magnetic field. Following \citet{Melrose1971}, the electron emissivity in the Stokes basis from Equations \ref{eq:S_equations} and \ref{eq:kernel} in the main text can be expressed in tensor form as
\begin{equation}
    \begin{split}
    \label{eq:jtensor}
        j^{\alpha \beta}& = \int d^3 p f(E, \xi) \eta^{\alpha \beta} = \int_0^{\infty} d E f(E, \theta_B) \frac{\sqrt{3}e^2 \nu_c \sin \theta_B}{8\pi c} H^{\alpha \beta}(X),\\
        &H^{11}(X) = X \left[ \int_{X}^\infty dt K_{5/3}(t) + K_{2/3}(X) \right],\\
        &H^{22}(X) = X \left[ \int_{X}^\infty dt K_{5/3}(t) - K_{2/3}(X) \right],\\
        &H^{12}(X) = -H^{21} = -\frac{2 i \cot \theta_B}{3 \gamma} \left[ (2 + g(\theta_B))\int_{X}^\infty dt K_{1/3}(t) + 2X K_{1/3}(X) \right],\\
    \end{split}
\end{equation}
where $\nu_c = eB/2\pi m_e c$ is a non-relativistic cyclotron frequency, $X = \nu / \nu_{cr}$ and $\nu_{cr} = (3/2) \nu_c \gamma^2 \sin \theta_B$.   The first expression in Equation~\ref{eq:jtensor} is general while in the second expression we have integrated over pitch angle $\xi$ by assuming $\xi \simeq \theta_B$. The Stokes emissivities are related to Equation~\ref{eq:jtensor} as $j_I = j^{22}+j^{11}$, $j_Q = j^{22}-j^{11}$, $j_U=j^{12}+j^{21} \equiv 0$, and $j_V = i(j^{12}-j^{21})$. Here, unlike in \citet{Melrose1971}, we define $g(\theta_B)$ for a general non-separable gyrotropic distribution function, which for our choice of the distribution (Equation~\ref{eq:df} in the main text) is
\begin{equation}
    g(\theta_B) = \tan \theta_B \left. \frac{df(E,\xi)}{d \xi} \right| _{\xi = \theta_B} \frac{1}{f(E,\theta_B)} = \frac{\gamma}{ \epsilon_\perp^\ast} \frac{(\eta-1) \sin^2 \theta_B}{ 1 + (\eta-1) \cos^2 \theta_B} = \frac{\gamma}{ \epsilon_\perp^\ast} \left(\frac{\epsilon_\perp^\ast}{ \epsilon_\perp}\right)^2 (\eta-1) \sin^2 \theta_B \equiv A \gamma.
    \label{eq:gtheta}
\end{equation}
In the last equality in Equation~\ref{eq:gtheta} we have defined the anisotropy parameter A (a function of $\eta)$ that will appear below.

We now proceed analytically evaluating the emissivities $j_I$, $j_Q$ and $j_V$, beginning with $j_I$. Equation \ref{eq:jtensor} for $j_I$ can be rewritten as:
\begin{equation}
    j_{I}(\epsilon_\perp, \nu/\nu_c, \theta_B, \eta) = \frac{\sqrt{3} B m_e^2 c e^3 \sin(\theta_B)}{8 \pi} \int d\gamma \gamma^2 \beta f(\gamma, \theta_B) F(X) = \eta^{1/2} \frac{\sqrt{3} n_e B e^3 \sin(\theta_B)}{32 \pi^2 m_e c^2 \epsilon_\perp K_2(\epsilon_\perp)} \int d\gamma \gamma^2 \beta e^{-\gamma/\epsilon_\perp^\ast} F(X),
    \label{eq:jIintermediate}
\end{equation}
where
\begin{equation}
    F(X) = X \int_{X}^\infty dt K_{5/3}(t) = 
    \begin{cases}
    2^{2/3}\Gamma(1/3) X^{1/3} + \mathcal{O}(X),\  X \ll 1,\\
    \sqrt{\frac{\pi}{2} X}  e^{-X} (1 + \mathcal{O}(1/X)),\  X \gg 1
    \label{eq:Fx}
    \end{cases}
\end{equation}
is the asymptotic behavior of the synchrotron power at low and high frequencies and $\Gamma(a)$ is the gamma-function. To express the emissivity in terms of the new temperature $\epsilon_\perp^\ast = \epsilon_\perp^\ast(\xi = \theta_B)=\epsilon_\perp/\sqrt{1 + (\eta - 1) \cos^2 \theta_B}$, as given by distribution in Equation~\ref{eq:df_high_gamma}, we consider separately the low and high frequency limits in Equation~\ref{eq:Fx} applied to Equation~\ref{eq:jIintermediate}. In the low-frequency limit, 
\begin{equation}
   j_I(\epsilon_\perp, \nu/\nu_c, \theta_B, \eta) \propto \eta^{1/2} \int_1^\infty d \gamma \gamma^2 \beta \frac{e^{-\gamma/\epsilon_\perp^\ast}}{\epsilon_\perp K_2(\epsilon_\perp)} \gamma^{-2/3} \approx \frac{\eta^{1/2} }{\epsilon_\perp K_2(1/\epsilon_\perp)} \int_1^{\infty} d\gamma \gamma^{4/3} e^{-\gamma/\epsilon_\perp^\ast} \approx \eta^{1/2}\frac{{\epsilon_{\perp}^\ast}^{7/3}}{\epsilon_\perp K_2(1/\epsilon_\perp)}.
\end{equation}
Therefore, the final expression for the emissivity is
\begin{equation}
    j_I(\epsilon_\perp, \nu/\nu_c, \theta_B, \eta) \approx \frac{2^{4/3} \pi \eta^{1/2}}{3} \frac{n_e e^2 \nu_s^\ast}{c {K_2(1/\epsilon_\perp)}} \left( \frac{\nu}{\nu_s^\ast} \right)^{1/3} \left( \frac{\epsilon_\perp^\ast}{\epsilon_\perp} \right) = \eta^{1/2} \frac{\epsilon_\perp^\ast}{\epsilon_\perp} \frac{K_2(1/\epsilon_\perp^\ast)}{K_2(1/\epsilon_\perp)} j_{I, {\rm iso}}(\epsilon=\epsilon_\perp^\ast, \nu/\nu_c, \theta_B),
    \label{eq:jIanalytic}
\end{equation}
where $\nu_s^\ast = \frac{2}{9} \nu_c \epsilon_\perp^{\ast 2} \sin \theta_B$ and $\epsilon_\perp^\ast = \epsilon_\perp^\ast(\xi=\theta_B)$. This calculation was done in the limit of low $\nu$, but the same expression can be obtained in the limit of high $\nu$ as well. The integral over Lorentz factor in Equation~\ref{eq:jIintermediate} now becomes
\begin{equation}
    j_I(\epsilon_\perp, \nu/\nu_c, \theta_B, \eta) \propto \eta^{1/2} \int_1^\infty d \gamma \gamma \beta \frac{e^{-\gamma/\epsilon_\perp^\ast - X}}{\epsilon_\perp K_2(1/\epsilon_\perp)}.
    \label{eq:jIint2}
\end{equation}
The maximum of the exponent in Equation~\ref{eq:jIint2} occurs at $\gamma_0 = (2 B \epsilon_\perp^\ast)^{1/3}$, where $B = (\nu / \nu_{cr}) \gamma^2 = 2/3 (\nu/\nu_c) \sin^{-1} \theta_B \gg 1 $.  The integral over $\gamma$ can then be carried out using the method of steepest descent (as in the case of an isotropic distribution function), leading again to 
\begin{equation}
    j_I(\epsilon_\perp, \nu/\nu_c, \theta_B, \eta)  = \eta^{1/2}\frac{\epsilon_\perp^\ast}{\epsilon_\perp} \frac{K_2(1/\epsilon_\perp^\ast)}{K_2(1/\epsilon_\perp)} j_{I, {\rm iso}}(\epsilon=\epsilon_\perp^\ast, \nu/\nu_c, \theta_B).
\end{equation}
The fact that $j_I$ for the anisotropic relativistic Maxwellian can be expressed as Equation~\ref{eq:jIanalytic} in both the low and high frequency limits motivates our using this expression as the proposed fit in Equation~\ref{eq:fits} of the main text. Physically, this corresponds to the total intensity emissivity just changing due to a different effective distribution function in the angle $\theta_B$ towards the observer. Note as well that although we derived Equation~\ref{eq:jIanalytic} for total intensity the same expression scaled to the isotropic distribution function emissivity holds for the intrinsic linear polarization emissivity, i.e., $j_Q$. This is because $K_{2/3}(X)$ has the same functional form as $\int_{X}^{\infty} K_{5/3}(X)$ at both high and low frequencies.

Circular polarization, however, has a different functional form:
\begin{equation}
\label{eq:jV}
    j_V(\epsilon_\perp, \nu/\nu_c, \theta_B, \eta) \propto \eta^{0.5} \cot \theta_B \sin \theta_B \int_1^{\infty} d \gamma \gamma \frac{e^{-\gamma/ \epsilon_\perp^\ast}}{\epsilon_\perp K_2(1/\epsilon_\perp)} \left[ (g(\gamma, \theta_B) + 2) \int_{X}^\infty K_{1/3}(t) dt + 2 X K_{1/3} \left( X \right)\right]. 
\end{equation}
Unlike in the case of total intensity and linear polarization, the circular polarization emissivity requires expanding the distribution function in a narrow cone around $\theta_B$; the resulting $j_V$ depends on the derivative of the distribution function, included in $g(\gamma,\theta_B)$. To understand the origin of our fit for $j_V$ in Equations~\ref{eq:fits}, we first consider the high-frequency limit when both $\int_{X}^\infty K_{1/3}(t) dt$ and $K_{1/3} \left( X \right)$ scale as $e^{-X}/ \sqrt{X}$ for $X \gg 1$. The integrand in equation \ref{eq:jV} can be written as $h(\gamma, A)e^{S(\gamma,A)}$, where 
\begin{equation}
    S(\gamma,A) = -\gamma/\epsilon_\perp^\star - B/\gamma^2 \ \ \ \ {\rm and} \ \ \ \ h(\gamma, A) = (A\gamma^3 + 2 \gamma^2 + 2B) \approx (A \gamma^3 + 2 B).
\end{equation}
The exponential term $e^{S(\gamma)}$ is again maximum at $\gamma_0 \approx (2B \epsilon_\perp^\ast)^{1/3}$, where $B = (\nu / \nu_{cr}) \gamma^2 = 2/3 (\nu/\nu_c) \sin^{-1} \theta_B \gg 1 $.  Equation~\ref{eq:jV} can then be integrated via the method of steepest descent as:
\begin{equation}
\label{eq:jVint}
    \int d \gamma h(\gamma) e^{S(\gamma)} \approx \sqrt{\frac{2 \pi}{-S''(\gamma_0)}} e^{S(\gamma_0)} h(\gamma_0) = \sqrt{\frac{2 \pi}{-S''(\gamma_0)}} e^{S(\gamma_0)} \times 2 B(A \epsilon_\perp^\ast + 1) = \sqrt{\frac{2 \pi}{-S''(\gamma_0)}} e^{S(\gamma_0)} \times 2 B \eta \left( \frac{\epsilon_\perp^\ast}{\epsilon_\perp}  \right)^2.
\end{equation}
As with $j_I$ and $j_Q$ we choose to express $j_V$ relative to the result for an isotropic Maxwellian with temperature $\epsilon^\star_\perp$. The latter can be derived in an identical manner to Equation~\ref{eq:jVint}. We find that ratio of $j_V$ in the anisotropic case to $j_{V, {\rm iso}}$ at a temperature of $\epsilon_\perp^\ast$ and $A=0$ has two terms.  One is the ratio of distribution function normalizations $\eta^{1/2} (\epsilon^\star_\perp/\epsilon_\perp) (K_2(1/\epsilon^\star_\perp)/K_2(1/\epsilon_\perp))$ that appears in $j_I$ and $j_Q$.   The other is the factor $\eta (\epsilon^\star_\perp/\epsilon_\perp)^2$ in Equation~\ref{eq:jVint} -- present only in $j_V$ and not $j_Q$ and $j_I$ -- that is due to the presence of the distribution function derivative $g(\theta_B)$ in the circular polarization emissivity.  The net result is
\begin{equation}
    \frac{j_V(\epsilon_\perp, \nu/\nu_c, \theta_B, \eta)}{j_{V, {\rm iso}}(\epsilon = \epsilon_\perp^\ast, \nu/\nu_c, \theta_B)} = \eta \left( \frac{\epsilon_\perp^\ast}{\epsilon_\perp}  \right)^2 \times \eta^{1/2} \left( \frac{\epsilon_\perp^\ast}{\epsilon_\perp}  \right) \left( \frac{K_2(1/\epsilon_\perp^\ast)}{K_2(1/\epsilon_\perp)} \right)= \eta^{3/2} \left( \frac{\epsilon_\perp^\ast}{\epsilon_\perp}  \right)^3 \left( \frac{K_2(1/\epsilon_\perp^\ast)}{K_2(1/\epsilon_\perp)} \right),
\end{equation}
which gives the analytical fit given by Equation~\ref{eq:fits} in the main text.  The same result can be derived in the low frequency limit via suitable expansion of Equation~\ref{eq:jV}.

\subsection{Comparison of Analytics and Numerics}
\begin{figure}
    \centering
    \includegraphics[width=\textwidth]{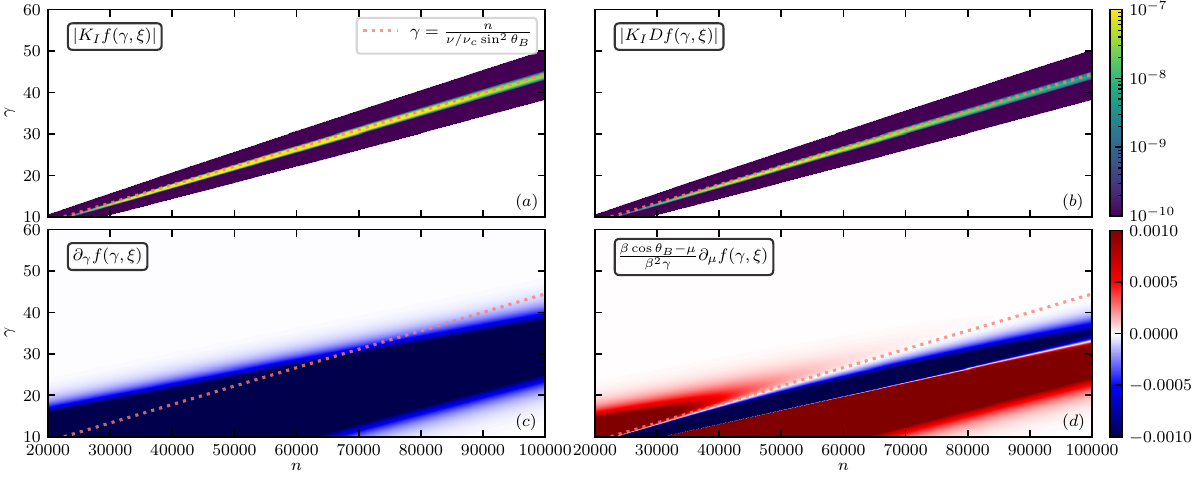}
    \caption{Integrands $|K_I f(\gamma, \xi)|$ for $j_I$ (a) and $|K_I Df(\gamma, \xi)|$ for $\alpha_I$ (b). Two parts of $Df$ that include $\partial_\gamma f(\gamma, \xi)$ (c) and $\partial_\mu f(\gamma, \xi)$ (d). The approximate location of the peak that corresponds to $\cos \theta_B = \beta \mu$ is shown by dotted lines. The free parameters are $\nu/\nu_c = 10^3$, $\eta=10$, $\epsilon_\perp=10$. While the location of the peak is still the same as in the $\eta=1$ case, a non-zero term with $\partial_\mu f(\gamma, \mu)$ appears which, however, goes through zero at the peak of the integrand. Note the saturated colorbar in (c) and (d).}
    \label{fig:kernel}
\end{figure}
We now solve Equations~\ref{eq:S_equations}-5 in the main text numerically and check the validity of the approximations used in the previous section for obtaining analytical fits for the polarized synchrotron emissivity and absorption coefficients.  To do so, we use the publicly available code {\tt symphony} to compare our theoretical fits (Eq.\ref{eq:fits}) with a numerical solution. We implemented an anisotropic distribution function to calculate $j_{S}$ and $\alpha_{S}$. In particular, we added the possibility for the distribution to depend on harmonic number $n$ as well as a non-zero $\partial_\mu f$ term in the absorption coefficient calculation (Eq.~\ref{eq:Df} that shows up in Eq.~\ref{eq:S_equations} includes $\partial_\mu f$) {-- both were absent }in {\tt symphony}. The distribution function and analytical derivatives $\partial_\gamma f$ and $\partial_\mu f$ can now depend on $\mu = \cos \xi$. However, as described in Section~\ref{sec:expressions} and below, the term with $\partial_\mu f$ in the absorption coefficient is negligible because it shows up proportional to a term that vanishes when the pitch angle is approximately $\theta_B$.  

The integrands in Equations~\ref{eq:S_equations} are $K_a f(\gamma, \xi)$ for $j_a$ and $K_a Df(\gamma, \xi)$ for $\alpha_a$, where $\xi$ can be substituted for $n$ since at $y_n=0$ (as required by $\delta(y_n)$):
\begin{equation}
    \cos \xi = \frac{1 - \frac{n}{\gamma}
    \frac{\nu_c}{\nu}}{\beta \cos \theta_B}.
\end{equation}
Thus, the integrands can be expressed as functions of $\gamma$ and $n$ and integrated in $\gamma-n$ space. 

In Figure~\ref{fig:kernel} we show the integrands for $j_I$ (a) and $\alpha_I$ (b) and the two terms from $Df$ that include $\partial_\gamma f(\gamma, \xi)$ and $\partial_\mu f(\gamma, \xi)$ for $\nu/\nu_c=10^3$, $\epsilon_\perp=10$, and $\eta=10$. The location of the sharp peak in the $\gamma-n$ plane is where $\xi \simeq \theta_B$ (as in the isotropic distribution function case).  However, the exact harmonic at which the emission peaks moves along this line depending on $\eta$. This is equivalent to the result in Figure~\ref{fig:jnu} that different $\theta_B$ dominate the emission as we vary $\eta$.  Panels (c) and (d) in Figure \ref{fig:kernel} show that, at the location in $\gamma-n$ space where the absorption coefficient peaks (panel b), the first term in $Df$ due to gradients in $\gamma$ is much larger than the second term due to gradients in $\mu$. This is because most of the emission and absorption is coming from pitch angles of $\xi \approx \theta_B$. Thus, the propagation is almost parallel, and the term with $\beta \cos \theta_B - \mu$ shown in panel (d) does not contribute significantly to $Df$.   This implies that in practice the total intensity emission and absorption coefficients for the anistropic distribution function are equivalent to calculations for a thermal isotropic distribution function at a new temperature $\epsilon^\star_\perp$.   This allows us to calculate $\alpha_a$ from $j_a$ via Kirchoff's law even for our anisotropic distribution function (at least in the limit of high $\gamma$ where $\xi \approx \theta_B$ is justified).

A number of fitting functions for $j_a$ and $\alpha_a$ are used in the literature [see, e.g., \citet{Pandya2016ApJ, Dexter2016MNRAS}]. Here we compare our results for the fits used by {\tt blacklight}:
\begin{equation} 
j_{S, {\rm iso}}(\epsilon, X, \theta_B)= \frac{n_e e^2 \nu_c}{c} e^{-X^{1/3}} \times 
\begin{cases}
\frac{\sqrt{2} \pi }{27} \sin(\theta_B) (X^{1/2} + 2^{11/12} X^{1/6})^2\ [S=I],\\
-\frac{\sqrt{2} \pi }{27} \sin(\theta_B) \Big( X^{1/2} + \Big( \frac{7\epsilon^{24/25} + 35}{10\epsilon^{24/25}+ 75} \Big)2^{11/12} X^{1/6} \Big)^2\ [S=Q],\\
0\ [S=U],\\
\frac{\cos \theta_B}{\epsilon} \Big( \frac{\pi}{3} + \frac{\pi}{3} X^{1/3}+ \frac{2}{300}X^{1/2} + \frac{2}{19} \pi X^{2/3} \Big)\ [S=V],
\end{cases}\label{eq:isotropic}
\end{equation}
where $X=\nu/\nu_s$. Absorption coefficients $\alpha_{S}$ for a thermal distribution can be obtained via Kirchoff's law. 

\begin{figure}
    \centering
    \includegraphics[width=0.99\textwidth]{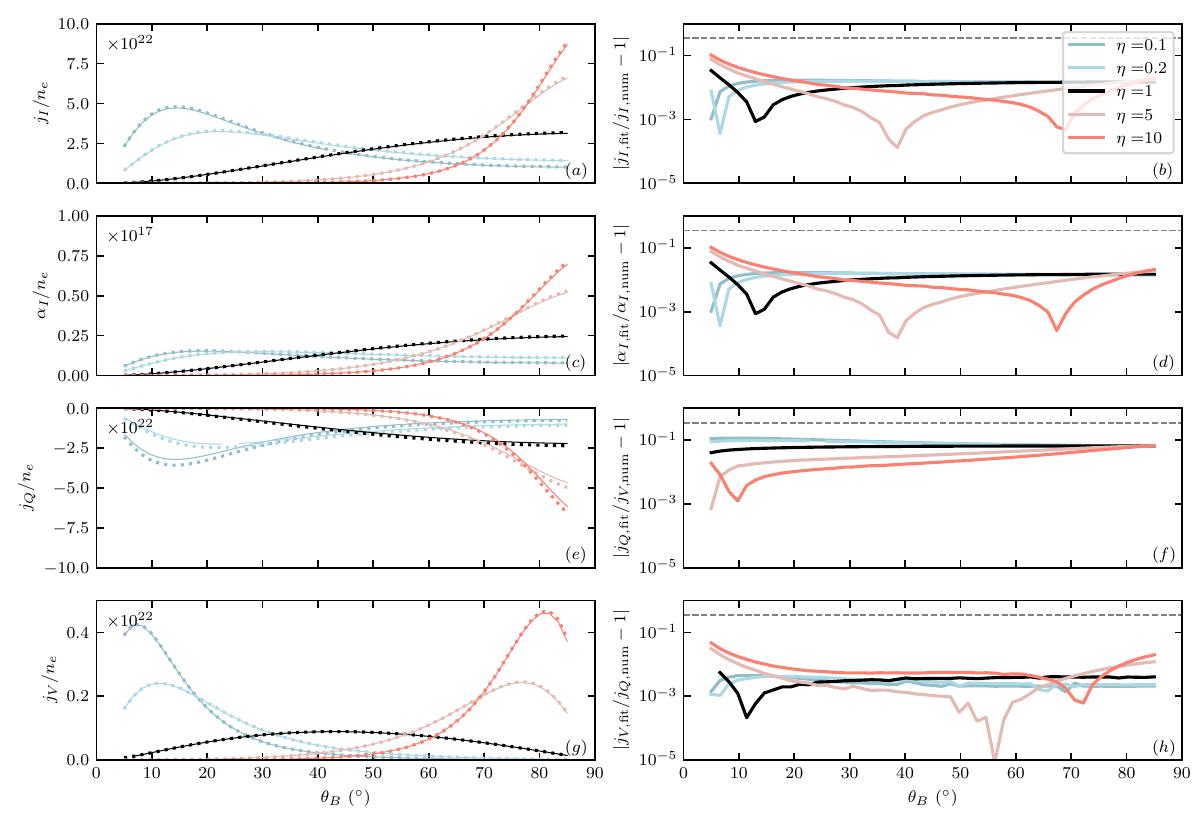}
    \caption{Comparison of numerical results for $j_{I}$ (a-b), $\alpha_{I}$ (c-d), $j_{Q}$ (e-f), and $j_V$ (g,h) with the theoretical fits given by Equations~\ref{eq:fits} and \ref{eq:isotropic}. Numerical results and theoretical fits are shown on the left by solid and dotted lines respectively, and on the right -- the relative error is shown. The dashed gray line on the right shows a relative error of $30\%$. The free parameters are $\epsilon_\perp=10$ and $\nu/\nu_c =10^3$.}
    \label{fig:numerical_fits}. 
\end{figure}
\begin{figure}
    \centering
    \includegraphics[width=0.99\textwidth]{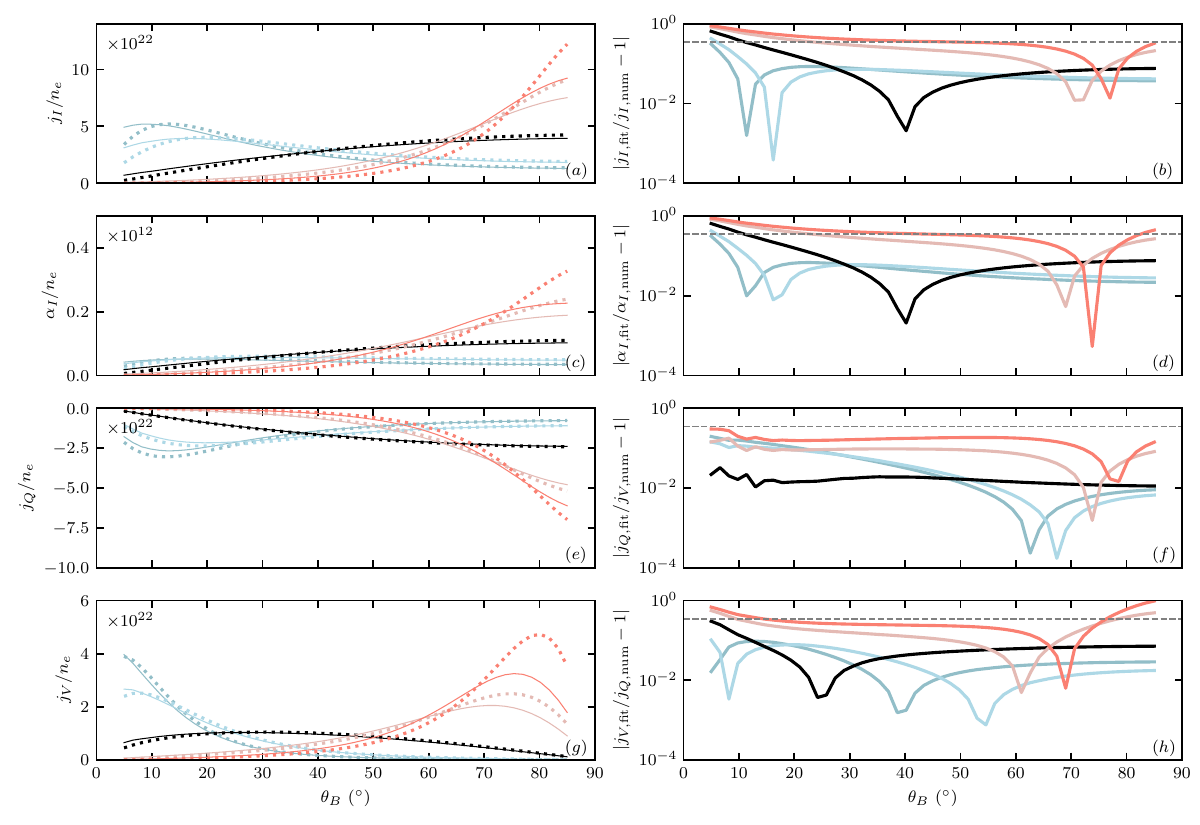}
    \caption{Same as Figure~\ref{fig:numerical_fits} but for $\epsilon_\perp=3$ and $\nu/\nu_c =10$.}
    \label{fig:numerical_fits_nr}. 
\end{figure}
Figure~\ref{fig:numerical_fits} shows numerical integration results from {\tt symphony} (solid lines) along with their respective theoretical fits (dotted lines) for $j_I$ (a), $\alpha_I$ (b), $j_Q$ (c), and $j_V$ (d) on the left. On the right, their respective absolute errors are shown in panels (b,d,f,h). All results are shown as a function of observing angle $\theta_B$ at different anisotropy parameters $\eta$ represented by different colors, at $\nu/\nu_c = 10^3$ and for $\epsilon_\perp = 10$.  These are typical parameters for application to Sgr A* and M87*.   The agreement is excellent for all $\eta$, with maximal errors $\lesssim 10\%$

We show a more challenging case of low temperature $\epsilon_\perp=3$ and low frequency $\nu / \nu_c = 10$ in Figure~\ref{fig:numerical_fits_nr}, which is organized identically to Fig.~\ref{fig:numerical_fits}.  This case is more challenging for our analytic fits than Figure~\ref{fig:numerical_fits} because the emission for $\epsilon_\perp=3$ and low frequency $\nu / \nu_c = 10$ is dominated by much lower energy electrons. The errors in our fits in Figure~\ref{fig:numerical_fits_nr} are, not surprisingly, larger. Generally, $\eta<1$ has smaller relative errors than $\eta>1$. This is because at a fixed observing angle $\theta_B$ and $\epsilon_\perp$, the effective temperature $\epsilon_\perp^\ast$ is larger than $\epsilon_\perp$ for $\eta < 1$. By contrast, $\epsilon_\perp^\ast<\epsilon_\perp$ for $\eta>1$, which can start to approach the non-relativistic cyclotron limit for which our fits do not apply. Figure~\ref{fig:numerical_fits_nr} shows that the case of $\eta<1$ has a relative error of $< 30 \%$ across all considered angles $\theta_B \in [5,85]^\circ$, and $< 10\%$ for most angles.  The fits have a relative error of larger than $30 \%$ for $\eta \geq 1$, however, at the largest and smallest angles.   This is true for the isotropic case as well at small angles.   We note, though, that the actual value of the emissivity and absorption coefficient are very small at small angles for $\eta \gtrsim 1$ (Fig.~\ref{fig:angle}) so that most of the emission and absorption will arise at larger angles where the fits are better. In addition at this low frequency, the emission will in most practical cases of interest be self-absorbed and approximately a blackbody. Finally, we note that a significant cause of error here is that our fits in Equation~\ref{eq:fits} for the anisotropic emission and absorption coefficients are scaled to the isotropic emissivity and absorption fits in Equation~\ref{eq:isotropic}, which become inaccurate at low temperatures, low frequencies, and small angles, as indicated by the large fractional errors for the isotropic case in Figure~\ref{fig:numerical_fits_nr}. In practice we advise caution in using the fits here if $\epsilon^\star_\perp \lesssim 3$ and the frequency is low $\lesssim 10 \nu_c$.  The regime of most interest for our applications is much higher frequencies where the analytic fits in Equation~\ref{eq:fits} are accurate.

\section{Anisotropy-driven instabilities in relativistic plasmas}
\label{ap:instability}
\subsection{Mirror instability}\label{ap:mirror}
To calculate the kinetic threshold for the relativistic mirror instability, we consider Vlasov and Maxwell's equations:
\begin{align}
    \frac{\partial f_s}{\partial t} + {\bf v}_s \cdot \nabla f_s + q_s ({\bf E} + \frac{{\bf v}_s}{c} \times {\bf B}) \cdot \frac{\partial f_s}{\partial {\bf p}}=0,\\
    \frac{1}{c}\frac{\partial {\bf E}}{ \partial t} = \nabla \times {\bf B} - \frac{4 \pi}{c} {\bf j}, \\
    \frac{1}{c}\frac{\partial {\bf B}}{ \partial t} = - \nabla \times {\bf E},
    \label{eq:vlasov}
\end{align}
where $s$ is the particle species (ions $i$ or electrons $e$) with mass $m_s$ and charge $q_s$, ${\bf v}_s = {\bf p}_s/ m_s \gamma$, ${\bf E}$ is electric field (with ${\bf E}_0 = 0$ initially), ${\bf B}$ is the magnetic field, and the axes are chosen such that ${\bf B} = B_0 \hat z$.  We now consider a small perturbation in the form of displacement $\propto e^{i {\bf kr}- i \omega t} $, where we consider ${\bf k } = k_\perp \hat x + k_\parallel \hat z$. We will initially consider electrons with an anisotropic distribution and ions with an isotropic distribution, $\delta {\bf E} = \delta E_y \hat y$ and thus $\delta {\bf B} = \delta B_x \hat x + \delta B_z \hat z$; the distribution function is perturbed as $f_s + \delta f_s$. The linearized equations are then
\begin{align}
    (-i \omega + i {\bf k} \cdot {\bf v}_s) \delta f_s + q_s \frac{{\bf v}_s}{c} \times {\bf B}_0 \cdot \partial_{\bf p} \delta f_s +  q_s \left( \delta {\bf E} +  \frac{{\bf v}_s}{c} \times \delta {\bf B} \right) \cdot \partial_{\bf p} f_s=0,\\
    \frac{4 \pi}{c^2} \omega \delta j_y = i \left(\frac{\omega^2}{c^2} - k^2 \right) \delta E_y, \ \frac{4 \pi}{c^2} \omega \delta j_x = i \left(\frac{\omega^2}{c^2} - k_\parallel^2 \right) \delta E_x,
\end{align}
where we used $\delta {\bf B} = \frac{c}{\omega} {\bf k } \times \delta {\bf E}$. We seek the solution of the linearized Vlasov equation for $\delta f_s$ and the corresponding current via the method of characteristics \citep[e.g.,][]{Mikhailovsky}. The current response $\delta j_y$ due to $\delta E_y$ is:
\begin{equation}
    \delta j_{y,s} = -2 \pi i q_s^2 \int d p d \mu p^2 \sum_{n=-\infty}^{+\infty} \frac{v^2 \sin^2 \xi}{\omega - k_\parallel v \cos \xi - n \Omega_s} \left[ \frac{1}{v} \frac{\partial f}{\partial p} - \frac{\cos \xi}{v p}\frac{\partial f}{\partial \mu} + \frac{k_\parallel}{\omega} \frac{1}{p} \frac{\partial f}{\partial \mu} \right] J^{'2}_n\left(\frac{k_\perp v_\perp}{\Omega_s}\right) \delta E_y,
    \label{eq:current_response}
\end{equation}
where $\Omega_s$ is the relativistic cyclotron frequency of species $s$. For $\Omega_s \gg \omega $ and $\Omega_s \gg k_\parallel v_\parallel$, keeping the leading terms $n=0, \pm1$ of order $\Omega_s^{-2}$ and using $J^{'}_0(z) \approx -\frac{z}{2}$ and $J^{'}_{\pm 1}(z) \approx \pm \frac{1}{2}$:
\begin{equation}
\begin{split}
    n=0: \ &-\frac{\pi i q_s^2}{2} \int d p d \mu p^2 \frac{v^4 \sin^4 \xi}{\omega - k_\parallel v \cos \xi} \left[ \frac{1}{v} \frac{\partial f_s}{\partial p} - \frac{\cos \xi}{v p}\frac{\partial f_s}{\partial \mu} + \frac{k_\parallel}{\omega} \frac{1}{p} \frac{\partial f_s}{\partial \mu} \right] \frac{k_\perp^2}{\Omega_s^2} \delta E_y, \\
    n=\pm1: \ &\frac{\pi i q_s^2}{2} \int d p d \mu p^2 v^2 \sin^2 \xi \frac{\omega - k_\parallel v \cos \xi}{\Omega_s^2} \left[ \frac{1}{v} \frac{\partial f_s}{\partial p} - \frac{\cos \xi}{v p}\frac{\partial f_s}{\partial \mu} + \frac{k_\parallel}{\omega} \frac{1}{p} \frac{\partial f_s}{\partial \mu} \right] \delta E_y.
\end{split}
\end{equation}

For isotropic ions the terms with $\partial_\mu f$ can be dropped, resulting in the following current response
\begin{equation}
    \delta j_{y,i} = \frac{\pi ic}{2B^2} \delta E_y \int d p d \mu p^3 \sqrt{m_i^2c^2 + p^2} S(\mu)\frac{\partial f_i}{\partial p} \left[ \omega - k_\parallel v \mu - \frac{v^2 S(\mu)}{\omega - k_\parallel v \mu} k_\perp^2 \right], 
    \label{eq:isotropic_ions}
\end{equation}
where the second term in brackets equals zero due to the odd function $S(\mu)=1- \mu^2 = \sin^2 \xi$. For the mirror mode, we are interested in the $k_\parallel v_\parallel / \omega \gg 1$ limit, which leaves only the resonant term. Using
\begin{equation}
    \frac{v^2}{\omega} \int_{-1}^{1} d \mu \frac{S^2(\mu)}{1 - \frac{k_\parallel v \mu}{\omega}} = -i\pi \frac{v}{k_\parallel} + \frac{16}{3} \frac{\omega}{k_\parallel^2} + \mathcal{O}((k_\parallel v_\parallel / \omega)^{-3}),
\end{equation}
and considering non-relativistic ions, $p/m_ic \ll 1$, with a Maxwellian distribution function $f_i(p)$ and number density $n$
\begin{equation}
    f_i(p) = \frac{n}{(2 \pi m_i^2 c^2 \epsilon_i)^{3/2}} e^{-p^2/2 m_i^2 c^2 \epsilon_i},
    \label{eq:ions_distribution}
\end{equation}
integration by parts of the third resonant term in Eq.~\ref{eq:isotropic_ions} results in
\begin{equation}
    \delta j_{y,i} = \frac{2 \pi^2 c^2}{B_0^2} \delta E_y \frac{k_\perp^2}{k_\parallel} \int_0^{\infty} dp p^3 f_s(p) = \frac{\delta E_y}{\sigma_i} \frac{c \epsilon_i^{1/2}}{4 (\pi/2)^{1/2}} \frac{k_\perp^2}{k_\parallel} = \frac{\delta E_y}{4 \pi \sigma_i }\pi^{1/2} v_{th,i} \frac{k_\perp^2}{k_\parallel},
\end{equation}
where $v_{th,i} = \sqrt{2 \epsilon_i}c$ and $\sigma_i = B_0^2 / 4 \pi n m_i c^2 = v_a^2 / c^2$, $v_a^2 = B_0^2 / 4 \pi n m_i$ is the Alfven speed.

We will now analyze the electron's current in Equation~\ref{eq:current_response}, splitting it by the three terms in the brackets $j_{y,e,1}$, $j_{y,e,2}$, and $j_{y,e,3}$. The second term in Eq.~\ref{eq:current_response} is $\mu \omega / k_\parallel v \ll 1$ times smaller than the third term and thus $j_{y,e,2}$ is negligible. As with the ions, considering the resonant term's residue of $- i \pi \omega/ k_\parallel v$ at $\mu_0 = \omega / k_\parallel v$
\begin{equation}
    \delta j_{y,e,1} \approx -\frac{\pi^2 c^2}{2 B^2} \delta E_y \frac{k_\perp^2}{k_\parallel} \int dp p^4 \frac{\partial f_e(p,\mu_0)}{\partial p} = \frac{\delta E_y}{4 \pi \sigma_i} \frac{2\pi^2}{m_i n} \frac{k_\perp^2}{k_\parallel} \int dp p^3 f_e(p, \mu_0).
\end{equation}

The dispersion relation in the limit of $\omega \ll k_\parallel v \mu$ is therefore
\begin{equation}
    -\omega \frac{k_\perp^2}{k_\parallel} \mathcal{J} = i k^2 v_a^2 + 4 \pi k_\parallel c \sigma_i \omega \frac{\delta j_{y,e,3}}{\delta B_x},
\end{equation}
where
\begin{equation}
    \mathcal{J} = \frac{2 \pi^2}{m_i n} \int dp p^3 f_e(p, \mu_0) + \pi^{1/2} v_{th,i} > 0.
\end{equation}

As in \citet{RelativisticMirror_Osipov_2017}, the current response from anisotropic electrons, which drives the mirror instability, can be calculated in the same form:
\begin{equation}
    \delta j_{y,e,3} = -i \frac{\pi c \delta B_x}{2 B_0^2} \frac{k_\perp^2}{k_\parallel} \int dp p^3 v \int_{-1}^{1} d \mu \frac{(1-\mu^2)^2}{\mu} \frac{\partial f_e(p,\mu)}{\partial \mu}, 
\end{equation}
which we will calculate in two parts
\begin{equation}
    \int_{-1}^{1} d \mu \frac{(1-\mu^2)^2}{\mu} \frac{\partial f_e}{\partial \mu} =  \underbrace{\int_{-1}^{1} d \mu (\mu^3 - 2\mu)\frac{\partial f_e}{\partial \mu}}_{I_1} +  \underbrace{\int_{-1}^{1} d \mu \frac{1}{\mu} \frac{\partial f_e}{\partial \mu}}_{I_2}.
\end{equation}

Integral $I_1$ can be calculated by parts and expressed through parallel and perpendicular pressure $P_{\parallel,e}$ and $P_{\perp,e}$ since
\begin{equation}
    I_1 = -2f_e(1) - \int_{-1}^1 f_e(p,\mu) d(\mu^3 - 2\mu) = -2f_e(1) + {2\int_{-1}^1 d\mu (1-\mu^2) f_e(p,\mu) - \int_{-1}^1 d\mu \mu^2 f_e(p, \mu)},
    \label{eq:I1_parts}
\end{equation}
and
\begin{equation}
    P_{\parallel,e} = 2\pi \int_0^{\infty} dp p^3 v \int_{-1}^1 d\mu \mu^2 f_e(p,\mu), \ P_{\perp,e} = \pi \int_0^{\infty} dp p^3 v \int_{-1}^1 d\mu (1-\mu^2) f_e(p,\mu).
\end{equation}
Therefore, the two integral terms in $I_1$ in Eq.~\ref{eq:I1_parts} lead to 
\begin{equation}
    2\pi \times 2 \int_0^{\infty} dp p^3 v \int_{-1}^1 d\mu (1-\mu^2) f_e(p,\mu) - 2\pi \int_0^{\infty} dp p^3 v \int_{-1}^1 d\mu \mu^2 f_e(p,\mu) = 4P_{\perp,e} - P_{\parallel,e} = P_{\parallel,e} \left( 4 \eta^\lambda - 1 \right).
\end{equation}

For the distribution function given by Equation~\ref{eq:df} in the main text, $f_e(p,\mu) \propto \exp{(-a \sqrt{1 + b\mu^2})}$, where $a=\gamma/\epsilon_\perp$ and $b=(1-1/\gamma^2) (\eta-1)$. Then, the boundary term in Eq.~\ref{eq:I1_parts} can be expressed as
\begin{equation}
    -2 \times 2\pi \int_0^{\infty} dp p^3 v f_e(p, \mu=1)
    = - \frac{n m_e c^2 \eta^{1/2} }{\epsilon_\perp K_2(1/\epsilon_\perp)}  \int_1^{\infty} d \gamma (\gamma^2 -1 )^{3/2} e^{-a \sqrt{1+b}} = - P_{\parallel,e} \frac{\eta^{1/2 + \lambda} }{\epsilon_\perp^2 K_2(1/\epsilon_\perp)}  \int_1^{\infty} d \gamma (\gamma^2 -1 )^{3/2} e^{-a \sqrt{1+b}},
\end{equation}
resulting in the following contribution to $\delta j_{y,e,3}$ from $I_1$:
\begin{equation}
    2\pi \int_0^{\infty} dp p^3 v I_1 = P_{\parallel,e} \left[ 4 \eta^\lambda -1 - \frac{\eta^{1/2 + \lambda} }{\epsilon_\perp^2 K_2(1/\epsilon_\perp)}  \int_1^{\infty} d \gamma (\gamma^2 -1 )^{3/2} e^{-a \sqrt{1+b}} \right].
\end{equation}

For calculating $I_2$, we find the derivative of the distribution function as $\partial f_e(p,\mu)/\partial \mu = -ab \frac{\mu}{\sqrt{1+b\mu^2}} f_e(p,\mu)$. Since the integrand is an even function of $\mu$, the contribution of $I_2$ to the current is
\begin{equation}
\begin{split}
        2 \pi \int_0^{\infty} dp p^3 v I_2 
        = - \frac{n m_e c^2 \eta^{1/2}}{\epsilon_\perp K_2(1/\epsilon_\perp)} \int_1^{\infty} d\gamma (\gamma^2-1)^{3/2} a b \int_0^{1} d\mu \frac{e^{-a\sqrt{1+b\mu^2}}}{\sqrt{1+b\mu^2}} \\
        = - P_{\parallel,e} \frac{ \eta^{1/2+\lambda}}{\epsilon_\perp^2 K_2(1/\epsilon_\perp)} \int_1^{\infty} d\gamma (\gamma^2-1)^{3/2} a b \int_0^{1} d\mu \frac{e^{-a\sqrt{1+b\mu^2}}}{\sqrt{1+b\mu^2}}.
\end{split}
\end{equation}
Therefore, the relevant current can be expressed as
\begin{equation}
\begin{split}
    \delta j_{y,e,3} &= -i \frac{c \delta B_x}{4 B_0^2} \frac{k_\perp^2}{k_\parallel} P_{\parallel,e}  \left[ 4 \eta^\lambda -1 - \frac{\eta^{1/2 + \lambda} }{\epsilon_\perp^2 K_2(1/\epsilon_\perp)} \left( \int_1^{\infty} d \gamma (\gamma^2 -1 )^{3/2} e^{-a \sqrt{1+b}} + \int_1^{\infty} d\gamma (\gamma^2-1)^{3/2} a b \int_0^{1} d\mu \frac{e^{-a\sqrt{1+b\mu^2}}}{\sqrt{1+b\mu^2}} \right) \right] \\
    & \equiv - i \frac{c \delta B_x}{4 B_0^2} \frac{k_\perp^2}{k_\parallel} P_{\parallel,e}  \left[ 4 \eta^\lambda -1 - \mathcal{I} \right] 
\end{split}
\end{equation}
which defines the integral $\mathcal{I}$.

Therefore, the final dispersion relation is
\begin{equation}
    -i \omega \frac{k_\perp^2}{k_\parallel} \mathcal{J} = k^2 v_{a,i}^2 
- k_\perp^2 \frac{P_{\parallel,e}}{4 \rho_i} (4 \eta^\lambda - 1 - \mathcal{I}),
\end{equation}
or, for $k_\perp \gg k_\parallel$
\begin{equation}
    -i\frac{\omega}{k_\parallel} \rho_i \mathcal{J} = \frac{B_0^2}{4 \pi} -  P_{\parallel,e}(4 \eta^\lambda - 1 - \mathcal{I})/4.
\end{equation}
The growth rate of the instability is positive when 
\begin{equation}
    2 \beta_{\parallel,e}^{-1} - (4 \eta^\lambda -1 - \mathcal{I})/4 > 0.
\end{equation}
Therefore, the threshold can be expressed as
\begin{equation}
    \beta_{\perp,e} < \frac{8 \eta^\lambda}{4 \eta^{\lambda} - 1 - \mathcal{I}},
    \label{eq:mirror_threshold}
\end{equation}
where
\begin{equation}
    \mathcal{I} = \frac{\eta^{1/2 + \lambda} }{\epsilon_\perp^2 K_2(1/\epsilon_\perp)} \left( \int_1^{\infty} d \gamma (\gamma^2 -1 )^{3/2} e^{-a \sqrt{1+b}} + \int_1^{\infty} d\gamma (\gamma^2-1)^{3/2} a b \int_0^{1} d\mu \frac{e^{-a\sqrt{1+b\mu^2}}}{\sqrt{1+b\mu^2}} \right).
    \label{eq:mirror_threshold_I}
\end{equation}
In the ultra-relativistic limit, $\sqrt{\gamma^2 - 1} \approx \gamma$ and $b \approx \eta-1$, $\mathcal{I}$ reduces to
\begin{equation}
\begin{split}
    \mathcal{I} &= \frac{\eta^{1/2 + \lambda} }{\epsilon_\perp^2 K_2(1/\epsilon_\perp)} \left( \int_1^{\infty} d \gamma \gamma^3 e^{-\frac{\gamma}{\epsilon_\perp} \sqrt{\eta}} + \frac{\eta - 1}{\epsilon_\perp} \int_0^{1} \frac{d\mu }{\sqrt{1+(\eta-1)\mu^2} }\int_1^{\infty} d\gamma \gamma^4 e^{-\frac{\gamma}{\epsilon_\perp}\sqrt{1+(\eta-1)\mu^2}} \right) \\ 
    & = \frac{3\eta^{1/2 + \lambda} \epsilon_\perp^2}{K_2(1/\epsilon_\perp)}  \left( \frac{2}{\eta^2}  + \left[ 3 - \frac{1}{\eta} - \frac{2}{\eta^2} + 3 \sqrt{\eta-1} \tan^{-1} \sqrt{\eta-1} \right]\right) \rightarrow \frac{9 \pi \eta^{1 + \lambda} \epsilon_\perp^2}{2 K_2(1/\epsilon_\perp)}, \ \eta \rightarrow \infty,
\end{split}
\end{equation}
where the boundary term  (first term in Eq.~\ref{eq:mirror_threshold_I}) cancels. Therefore, the mirror instability threshold in the ultra-relativistic limit is
\begin{equation}
    \beta_{\parallel,e} < \frac{8}{4 \eta^\lambda - 1 - 1.5\eta^{1/2 + \lambda}\left[ 3 + 3 \sqrt{\eta-1} \tan^{-1} \sqrt{\eta-1}  - 1/\eta \right]}.
    \label{eq:mirror_ur}
\end{equation}

In the non-relativistic limit, the instability threshold defined by Equations~\ref{eq:mirror_threshold} and \ref{eq:mirror_threshold_I} reduces to $\beta_{\perp,e} < 1/(\eta - 1)$, consistent with previous work. This is because for a non-relativistic distribution $P_{\perp,e}/P_{\parallel,e} = \eta^1$, i.e., $\lambda = 1$. The contribution to $\delta j_{y,e,3}$ from the boundary term in $I_1$ (first term in Eq.~\ref{eq:mirror_threshold_I}) is $3 P_{\perp,e}/\eta^2$ and the contribution from $I_2$ (second term in Eq.~\ref{eq:mirror_threshold_I}) is $P_{\perp,e}(\eta-1) (8 \eta^2 + 4 \eta +3) / \eta^2$. Thus, $4 \eta - 1 - \mathcal{I} = 8 \eta (\eta - 1)$ in a non-relativistic limit. 

In the case of non-relativistic anisotropic ions with anisotropy parameter $\eta_i$, the same derivation leads to a threshold
\begin{equation}
    \beta_{\perp,e} < \frac{8 \eta^\lambda}{4 \eta^{\lambda} - 1 - \mathcal{I} + 8 \frac{P_{\parallel,i}}{P_{\parallel,e}}\eta_i(\eta_i-1)}.
\end{equation}
Therefore, the threshold condition is defined by the ions as $\beta_{\perp,i} < 1/(\eta_i -1 )$ when $P_{\parallel,i} \gg P_{\parallel,e}$.

The assumption of a zero parallel current $j_z$ holds only when plasma-$\beta$ of at least one species is $\ll 1$. Consequently, when this assumption is invalid, a non-zero $\delta E_z$ also leads to a more complex $j_y$. In this case, both $j_y$ and $j_z$ contain terms proportional to $\delta E_y$ and $\delta E_z$ through Vlasov's equation. In a non-relativistic limit, the threshold is modified by an additional stabilizing term, which depends on plasma-$\beta$ of all species [see, e.g., \citet{hall_1979, Hellinger2007}]. A similar stabilizing relation can be obtained in a relativistic limit. The dispersion relation in the long-wavelength limit with $\Omega_e \gg \omega$ and $\Omega_e \gg k v$ is then defined by the following two inseparable equations:
\begin{equation}
    \frac{4 \pi}{c^2} \omega \delta j_y = i \left( \frac{\omega^2}{c^2} - k^2 \right) \delta E_{y}, \ \frac{4 \pi}{c^2} \omega \delta j_z = i \left( \frac{\omega^2}{c^2} - k_\perp^2 \right) \delta E_{z},
\end{equation}
which is equivalent to writing it in terms of plasma dielectric tensor $\mathcal{E}_{\alpha \beta}$:
\begin{equation}
    \mathcal{E}_{22} - \frac{\mathcal{E}_{23} \mathcal{E}_{32}}{\mathcal{E}_{33} - \frac{k_\perp^2 c^2}{\omega^2}} = \frac{k^2 c^2}{\omega^2},
\end{equation}
where $\mathcal{E}_{22}$ has already been calculated as the current response along $\hat y$ due to $\delta E_y$:
\begin{equation}
    \mathcal{E}_{22} = 1 - \frac{i k_\perp^2}{\omega k_\parallel^2 \sigma_i} \mathcal{J} + \frac{\pi c^2 k_\perp^2}{B_0^2 \omega^2} P_{\parallel,e} (4 \eta - 1 - \mathcal{I}).
\end{equation}

The general relations for relevant dielectric tensor components are:
\begin{equation}
\begin{split}
    \mathcal{E}^s_{23} &= -\mathcal{E}^s_{32} = \frac{4 \pi i q_s^2}{\omega} \int p_\perp v_\perp v_\parallel d p_\perp d p_\parallel \sum_{n=-\infty}^{+\infty} \frac{1}{\omega - k_\parallel v_\parallel - n \Omega_s} \left[\frac{1}{v_\perp} \frac{\partial f_s}{\partial p_\perp} + \frac{k_\parallel}{\omega} \left( \frac{\partial f_s}{\partial p_\parallel} - \frac{v_\parallel}{v_\perp} \frac{\partial f_s}{\partial p_\perp} \right) \right] J_n \left( \frac{k_\perp v_\perp}{\Omega_s} \right) J'_n \left( \frac{k_\perp v_\perp}{\Omega_s} \right) \\
    \mathcal{E}^s_{33} &= \frac{4 \pi q_s^2}{\omega} \int p_\perp v_\parallel d p_\perp d p_\parallel \sum_{n=-\infty}^{+\infty} \frac{1}{\omega - k_\parallel v_\parallel - n \Omega_s} \left[ \frac{\partial f_s}{\partial p_\parallel} - \frac{n \Omega_s}{\omega} \left( \frac{\partial f_s}{\partial p_\parallel} - \frac{v_\parallel}{v_\perp} \frac{\partial f_s}{\partial p_\perp} \right) \right] J_n^2\left( \frac{k_\perp v_\perp}{\Omega_s} \right),
\end{split}
\end{equation}
where $\mathcal{E}_{\alpha \beta} = \delta_{\alpha \beta} + \sum_s \mathcal{E}^s_{\alpha \beta}$. Keeping the leading terms of order $\Omega_s^{-1}$ for $\delta j_{z,\delta E_y}$ and $j_{y,\delta E_z}$ and terms of order $\Omega_s^0$ for $\delta j_{z,\delta E_z}$
\begin{equation}
\begin{split}
    \mathcal{E}^s_{23} = -\mathcal{E}^s_{32} &= 2 i\pi\frac{k_\perp}{\omega^2} \frac{q_s c m_s}{B_0} \int d p_\perp d p_\parallel \frac{v_\parallel p^2_\perp}{\omega - k_\parallel v_\parallel} \left[ k_\parallel v_\perp \frac{\partial f_s}{\partial p_\parallel} + (\omega - k_\parallel v_\parallel) \frac{\partial f_s}{\partial p_\perp} \right] \\ 
    &= - 2i \pi\frac{k_\perp k_ \parallel}{\omega^2} \frac{q_s c m_s}{B_0} \int d p_\perp d p_\parallel p_\perp^2 f_s \frac{\partial}{\partial p_\parallel} \left( \frac{v_\parallel v_\perp}{\omega - k_\parallel v_\parallel} \right) \\
    & = - 2 i\pi\frac{k_\perp}{\omega m_s k_ \parallel} \frac{q_s c}{B_0} \int d p_\perp d p_\parallel \frac{f_s p_\perp^3}{\gamma^4 (\omega/k_\parallel -  v_\parallel )^2} \left( \gamma^2 - 2 \frac{p_\parallel^2}{m_s^2 c^2} + \frac{k_\parallel v_\parallel}{\omega} \frac{p_\parallel^2}{m_s^2 c^2} \right), \\
    \mathcal{E}^s_{33} &= \frac{4 \pi q_s^2}{\omega} \int d p_\perp d p_\parallel \frac{p_\perp v_\parallel}{\omega - k_\parallel v_\parallel} \frac{\partial f_s}{\partial p_\parallel} = -\frac{4 \pi e^2}{m_s} \int d p_\perp d p_\parallel \frac{p_\perp f_s}{\gamma^2 (\omega - k_\parallel v_\parallel)^2} \left(\gamma - p_\parallel \frac{\partial \gamma}{\partial p_\parallel} \right) \\
    &= -\frac{4 \pi q_s^2}{k_\parallel^2 m_s} \int d p_\perp d p_\parallel \frac{p_\perp f_s}{\gamma^3 (\omega/k_\parallel - v_\parallel)^2} \left(1 + \frac{p_\perp^2}{m^2 c^2}\right).
\end{split}
\end{equation}
In a non-relativistic limit for an isotropic distribution, this reduces to:
\begin{equation}
\begin{split}
    \mathcal{E}^s_{23} &= - \frac{2 \pi k_\perp}{\omega  k_ \parallel} \frac{q_s c}{B_0} \int \frac{d p_\perp d p_\parallel p_\perp^2 v_\perp f_s}{(\omega/k_\parallel - v_\parallel)^2}, \\
    \mathcal{E}^s_{33} &= -\frac{4 \pi e^2}{m_s k_\parallel^2} \int \frac{d p_\perp d p_\parallel p_\perp f_s}{(\omega/k_\parallel - v_\parallel)^2} \approx \frac{\omega_{p,s}^2}{v_{th,s}^2 k_\parallel^2} \left( 1 + i \sqrt{\frac{\pi}{2}}\frac{\omega}{k_\parallel v_{th,s}} \right).
\end{split}
\end{equation}
In the dispersion relation, the imaginary term in $\mathcal{E}_{33}^s$ above will group with other imaginary terms in $\mathcal{E}_{22}$ as a coefficient in front of the growth rate $-i \omega$. Thus, the threshold condition, considering $k_\parallel v_\parallel / \omega \gg 1$ and $k_\perp \gg k_\parallel$, to leading order in $\Omega_s$, is
\begin{equation}
    \beta_{e,\parallel} \frac{4 \eta - 1 - \mathcal{I}}{8} + \beta_{\parallel,i} \eta_i (\eta_i - 1) > 1 - \frac{\pi}{B_0^2} \frac{\left( \sum_s \frac{q_s}{m_s} \int p_\perp dp_\perp dp_\parallel \frac{f_s p_\perp^2 (\gamma^2 - 2p_\parallel^2/m_s^2 c^2)}{\gamma^4 (\omega/k_\parallel - v_\parallel)^2} \right)^2}{ \sum_s \frac{q_s^2}{m_s} \int p_\perp d p_\perp dp_\parallel \frac{f_s (1 + p_\perp^2/m_s^2 c^2)}{\gamma^3 (\omega / k_\parallel - v_\parallel)^2}}.
\end{equation}

\begin{figure}
    \centering
    \includegraphics[width=\textwidth]{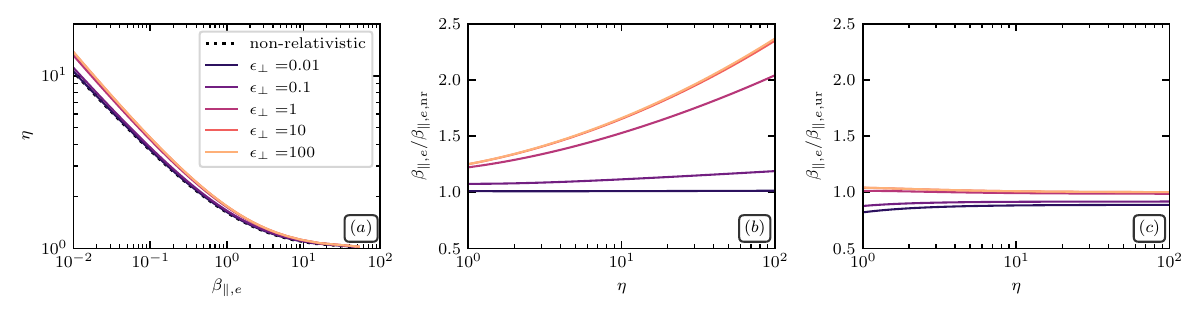}
    \caption{(a): Relativistic mirror instability thresholds (anisotropy $\eta$ as a function of $\beta_{\parallel,e}$) defined by Equations~\ref{eq:mirror_threshold}-\ref{eq:mirror_threshold_I} calculated for different $\epsilon_\perp$ represented by different colors from darkest ($\epsilon_\perp=0.01$) to brightest ($\epsilon_\perp = 100$). Dotted black line represents a non-relativistic limit. (b): Ratio of relativistic mirror instability threshold value $\beta_{\parallel,e}$ and its non-relativistic limit $\beta_{\parallel,e, {\rm nr}}$ as a function of $\eta$. (c): Ratio of relativistic mirror instability threshold value $\beta_{\parallel,e}$ and its ultra-relativistic limit $\beta_{\parallel,e, {\rm ur}}$, defined by Equation~\ref{eq:mirror_ur} as a function of $\eta$.}
    \label{fig:mirror}
\end{figure}
In Figure~\ref{fig:mirror} we show a comparison of the non-relativistic mirror threshold $\beta_{\parallel,e, {\rm nr}} = 1/(\eta-1)$ (thick black line) and the numerically calculated relativistic electron mirror threshold from Equations~\ref{eq:mirror_threshold} and \ref{eq:mirror_threshold_I}. In panel (a), $\eta$ is shown as a function of $\beta_{\parallel,e}$ at different temperatures. Panel (b) shows the ratio of the relativistic threshold value $\beta_{\parallel,e}$ and the non-relativistic threshold $\beta_{\parallel,e,{\rm nr}}$ as a function of $\eta$. Deviations are small for $\epsilon_\perp \lesssim 0.1$ and small values of anisotropy parameter $\eta \lesssim 10$. At high temperatures, the numerical solution is well approximated by the ultra-relativistic limit $\beta_{\parallel,e, {\rm ur}}$ given by Equation~\ref{eq:mirror_ur}. We show their ratio, $\beta_{\parallel,e, {\rm ur}}/\beta_{\parallel,e} $, as a function of $\eta$ in panel (c). Due to the large uncertainties in the electron anisotropy in accretion flows and since we limit the anisotropy $T_{\parallel,e}/T_{\perp,e}$ to be $\leq 10$ where the non-relativistic and relativistic mirror thresholds are similar, we chose to use the analytically simple non-relativistic mirror threshold for our application to BH images in \S~\ref{sec:imaging}.

\subsection{Parallel firehose instability}\label{ap:firehose}
To calculate the relativistic firehose threshold, we linearize Equations~\ref{eq:vlasov} for ${\bf k} = k \hat{z}$, $\delta {\bf B} = \delta B_y \hat y$, and $\delta {\bf E} = \delta E_x \hat x$ [see, e.g., \citet{RelativisticFirehose1973ApJ}]. The resulting equations for the ion and relativistic electron's currents are:
\begin{equation}
\begin{split}
    \delta j_{x,e} = -\frac{i \pi \delta E_x}{B_0^2 c^2} \int dp p^3 v d \mu S(\mu) (k c \mu - \omega) \left(p \frac{\partial f_e}{\partial p} - \mu \frac{\partial f_e}{\partial \mu} + \frac{k c}{\omega} \frac{\partial f_e}{\partial \mu} \right),\\
    \delta j_{x,i} = -\frac{i \pi \delta E_x}{B_0^2} \int dp d \mu p^3\sqrt{m_i^2 + (p/c)^2} S(\mu) (k v \mu - \omega) \frac{\partial f_i}{\partial p} = -4 i \omega \pi m_i c^2 \frac{\delta E_x}{B_0^2} \int dp p^2 \frac{4(p/m_ic)^2/3 + 1}{\sqrt{(p/m_ic)^2 + 1}} f_i(p).
\end{split}
\end{equation}
Solving for sub-relativistic isotropic ions with distribution function~\ref{eq:ions_distribution} and dropping the terms with odd $\mu$-integrands in $\delta j_{x,e}$ we find:
\begin{equation}
    \delta j_x = -\frac{i\omega \delta E_x}{4 \pi \sigma_i} \left[1 + \frac{5}{2} \epsilon_i + \frac{\pi }{n m_i c} \int dp d\mu p^3 (3- \mu^2)f_e(p,\mu)  + \frac{1}{n m_i c^2} (P_{\parallel,e} - P_{\perp, e})\frac{c^2 k^2}{\omega^2} \right].
\end{equation}
This results in the following dispersion relation
\begin{equation}
\begin{split}
    \omega^2 = k^2 c^2 \frac{\sigma_i - (P_{\parallel,e} - P_{\perp,e})/(n m_i c^2)}{\mathcal{F}},\\
    \mathcal{F} = \sigma_i + 1 + \frac{5}{2} \epsilon_i + \frac{2 \pi}{n m_i c} \int dp d \mu p^3 f_e(p,\mu) + \frac{P_{\perp,e}}{n m_i c^2}  >0,
\end{split}
\end{equation}
which gives the usual non-relativistic firehose threshold
\begin{equation}
    \frac{P_{\perp,e}}{P_{\parallel,e}} < 1 - \frac{2}{\beta_{\parallel,e}},
\end{equation}
where $\beta_{\parallel,e} = 8 \pi P_{\parallel,e} / B_0^2$. Note that if the ions are anisotropic as well the relevant firehose threshold becomes $P_{\perp,e} + P_{\perp,i} < P_{\parallel,e} + P_{\parallel,i} - B^2/4\pi$.  Thus if $P_{\perp,i} > P_{\perp,e}$ the ion anisotropy will in general be more important than the electron anisotropy in setting stability to the fluid firehose instability.

The calculation presented here focuses on the fluid parallel firehose instability. There are also resonant parallel and oblique firehose instabilities: Larmor-scale resonant instabilities destabilized by cyclotron interaction. The resonant instabilities typically have faster growth rates and somewhat lower anisotropy thresholds than the fluid firehose instability \citep{Gary1998,Hellinger2000}. Calculations of electron-scale resonant firehose instabilities for relativistically hot electrons with $T_p \gtrsim T_e$ would be valuable but we leave this to future work.

\subsection{Whistler instability}\label{ap:whistler}
The electron whistler instability, first noted in \citet{NonrelWhistlerSudan} and followed by a relativistic derivation \citep{RelWhistlerSudan}, is an instability of circularly polarized electron waves propagating along the magnetic field direction $B_0 \hat z$. Considering a wavevector $k$ and fluctuating electric field $\delta E_x$ and $\delta E_y$, the dispersion relation can be written as \citep{RelativisticWhistlerGlad1983}
\begin{equation}
    \frac{\epsilon_\parallel  k^2 c^2}{\beta_{\parallel,e}} - \frac{\epsilon_\perp \omega^2}{\beta_{\perp,e}} + \pi m_e^2 c^4 \Omega_{e,0}^2 \int \frac{p_\perp^2 v_\perp dp_\perp dp_\parallel}{kv_\parallel - (\omega - \Omega_e)} \left[ \frac{\partial f_e}{\partial p_\perp^2} (\omega - kv_\parallel) + \frac{\partial f}{\partial p_\parallel^2} kv_\parallel \right]=0,
\end{equation}
where $\epsilon_\parallel = T_{\parallel,e}/m_e c^2$, $f_e$ is defined by Equation~\ref{eq:df}, $p_\parallel$ and $p_\perp$ are relativistic parallel and perpendicular momentum, respectively, $\Omega_e$ and $\Omega_{e,0}$ are the relativistic and non-relativistic electron cyclotron frequencies. The whistler instability, like the ion cyclotron instability and unlike the mirror and firehose instabilities considered in the previous section, typically does not have a formal threshold but the growth rate becomes negligible for decreasing anisotropy.  This dispersion relation can thus be solved numerically to find the target growth rate for a fixed $\beta_{\perp,e}$ and varying $\eta$. 

The threshold for the relativistic whistler instability can be parameterized as \citep{Lynn2014} $P_{\perp,e} /P_{\parallel,e} = 1 + S(\epsilon_\perp)/\beta_{\perp,e}^\alpha$, where $S(\epsilon_\perp) = 0.265 - 0.165(1+\epsilon_\perp^{-1})$ and $\alpha = 0.58 - 0.043 \log \Gamma$ , where $\Gamma \sim 10^{-6} |eB/m_e c|$ is the assumed growth rate. Since $S(\epsilon_\perp)$ is a slowly varying and monotonic function of temperature, $S(\epsilon_\perp)\approx [0.1 - 0.25]$ for $\epsilon_\perp=[10^{-2},10^2]$, we choose to use $S(\epsilon_\perp=1) = 0.183$.

\section{Anisotropy model for GR radiative transfer}\label{ap:model}
\begin{figure}
    \centering
    \includegraphics[width=0.485\textwidth]{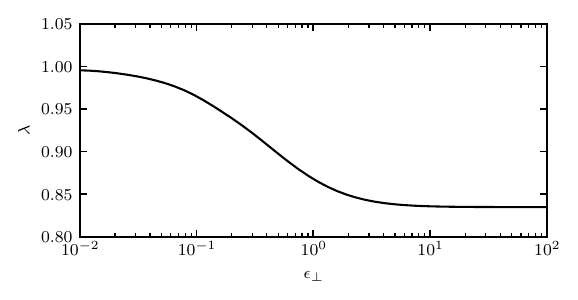}
    \caption{Numerically calculated $\lambda$ as a function of $\epsilon_\perp$ for $T_{\perp,e}/T_{\parallel,e} = \eta^\lambda$ for an anisotropic relativistic bi-Maxwellian distribution function. A simple fit for $\lambda$ is given in Equation~\ref{eq:lambdafit}. }
    \label{fig:eta_fit}
\end{figure}
For our choice of the distribution function (Eq.~\ref{eq:df}), the ratio of perpendicular and parallel temperatures $T_{\perp,e}/T_{\parallel,e} = \eta^\lambda$. The value of $\lambda$ is in turn a function of temperature $\epsilon_\perp$, which we show in Figure~\ref{fig:eta_fit}. The function $\lambda(\epsilon_\perp$) can be well-approximated by
\begin{equation}
\label{eq:lambdafit}
    \lambda = -0.08\tanh{\left( 1.5(\log_{10} \epsilon_\perp+0.5) \right)} + 0.92,
\end{equation}
In the non-relativistic limit, when $\epsilon_\perp \ll 1$, this gives $T_{\perp,e}/T_{\parallel,e} \approx \eta$, while in the ultra-relativistic limit, $T_{\perp,e}/T_{\parallel,e} \approx \eta^{0.8}$.

In our modeling of black hole accretion images we consider three limiting cases for the anisotropy of the distribution function $T_{\perp,e}/T_{\parallel,e}$, intended to bracket the magnitude of the effect that an anisotropic distribution function can introduce:
\begin{equation}
\begin{split}
    & \left( T_{\perp,e}/T_{\parallel,e} \right)_{\rm mirror}  = 1+1/\beta_{\perp,e},\\
    & \left (T_{\perp,e}/T_{\parallel,e} \right)_{\rm whistler} = 1+ S/\beta_{\perp,e}^\alpha, \\
    & \left (T_{\perp,e}/T_{\parallel,e} \right)_{\rm isotropic} \equiv 1, \\
    & \left (T_{\perp,e}/T_{\parallel,e} \right)_{\rm firehose} = 1-2/\beta_{\parallel,e},
\end{split}\label{eq:thresholds}
\end{equation}
where we take $S = 0.183$ and $\alpha = 0.838$ as in Appendix \S \ref{ap:whistler}. Since the firehose threshold is undefined at small electron-$\beta_{\parallel,e}$, we choose to set the threshold to a constant value of $T_{\perp,e}/T_{\parallel,e}=0.1$ 
at low $\beta_{\parallel,e}$. This is motivated by local simulations \citep{Riquelme2015ApJ}. Likewise, for the mirror instability, we limit $T_{\perp,e}/T_{\parallel,e} < 10$.   In reality, the temperature anisotropy at low $\beta$ will depend on the heating, expansion and contraction of the plasma, which is what drives the temperature anisotropy in the first place.

\begin{figure}
    \centering
    \includegraphics[width=0.5\columnwidth]{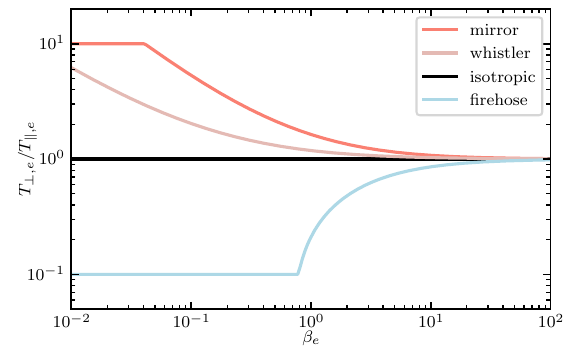}
    \caption{Models used in the radiative transfer of GRMHD simulation, represented by different colors: mirror and whistler ($T_{\perp,e}>T_{\parallel,e}$) and firehose ($T_{\perp,e} < T_{\parallel,e}$).}
    \label{fig:eta_models}
\end{figure}

It is useful to re-express Equations~\ref{eq:thresholds} in terms of the total electron temperature
\begin{equation}
    T_e = \frac{1}{3} (2T_{\perp,e} + T_{\parallel,e}). 
\label{eq:temperature_e}
\end{equation}
Using Equation~\ref{eq:temperature_e}, Equations~\ref{eq:thresholds} for the instability thresholds can be rewritten as
\begin{equation}
\begin{split}
    \beta_{\perp, e,{\rm mirror}} & = \frac{\beta_e}{2} - \frac{1}{3} + \frac{1}{2} \sqrt{\frac{4}{9} + \frac{8}{3}\beta_e + \beta_e^2},\\
    \beta_{\parallel, e,{\rm firehose}} & = \beta_e + \frac{4}{3},\\
    \beta_{\perp, e, {\rm whistler}} & = 
    \begin{cases}
        0.141 \beta_e^3 - 0.26 \beta_e^2 + 1.171 \beta_e + 0.005, \beta_e < 1 \\
        1.054 \beta_e + 0.012, \beta_e > 1\\
    \end{cases},
\end{split}
\label{eq:thresholds_betae}
\end{equation}
where $\beta_{\parallel, e,{\rm firehose}}$ and $\beta_{\parallel,e, {\rm mirror}}$ are exact solutions and $\beta_{\perp, e,{\rm whistler}}$ is a polynomial fit to a numerical solution with growth rate $\Gamma \sim 10^{-6} |eB/m_e c|$  and $\epsilon_\perp=1$. The thresholds in Equation~\ref{eq:thresholds_betae} can then be used in Equation~\ref{eq:thresholds}, thus providing expressions for the threshold temperature anisotropy in terms of $\beta_e$. This is a variable accessible to a simulation that does not evolve temperature anisotropy, such as those that we used in \S~\ref{sec:imaging}. The threshold conditions are shown in Fig.~\ref{fig:eta_models} as a function of electron $\beta_e$, which is extracted from the MHD plasma-$\beta_{\rm th}$ via $\beta_e = 2\beta_{\rm th}/(R+1)$.

j\section{GRMHD simulations}\label{ap:simulations}
The GRMHD simulations used in \S \ref{sec:imaging} were performed using the publicly available code {\tt Athena++} in spherical Kerr-Schild coordinates with a logarithmically stretched grid in the radial direction $r$. The setup is identical to \cite{White2019ApJMAD} with the outer radius of $1000r_g$ and the inner radius being inside the horizon. The grid is refined with the level $0$ grid being $N_r \times N_\xi \times N_\phi = 64 \times 32 \times 64$. A total of 3 refinement levels are concentrated around the midplane, $\theta=\pi/2$, resulting in an effective resolution of  $512 \times 256 \times 512$ in $r$, $\theta$, and $\phi$. We initialize a Fishbone torus \cite{1976ApJtorus} with a purely poloidal magnetic field with mean plasma-$\beta$ of $100$. We study two different spin values of the BH: $a=0.98$ and $0.5$. Each of the two simulations is run up to a steady state and several eruption events for a total simulation time of more than $15000r_g/c$.

\begin{figure*}
    \centering
    \includegraphics[width=0.99\columnwidth]{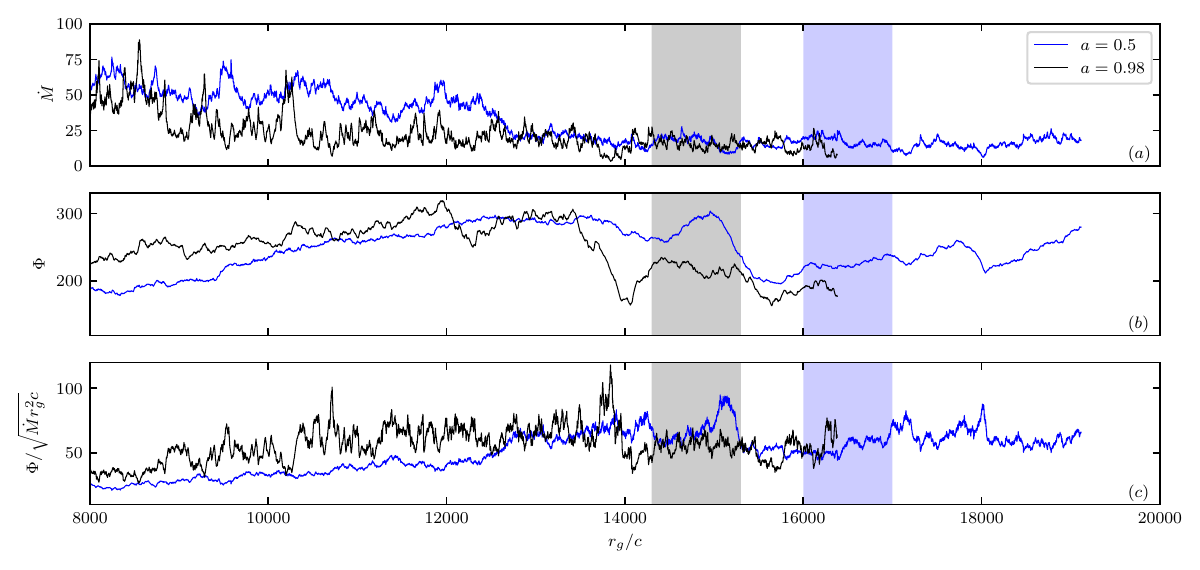}
    \caption{Time evolution of accretion rate $\dot{M}$ (a), magnetic flux $\Phi$ (b), and normalized magnetic flux $\Phi_{\rm BH} / \sqrt{\dot M r_g^2 c}$ after $8000r_g/c$ measured at $2 r_g$ for two spin parameters: $a=0.98$ (black) and $a=0.5$ (blue). Shaded regions correspond to quiescent periods of $1000r_g/c$ chosen for GR radiative transfer analysis: $14300-15300 r_g/c$ for $a=0.98$ and $16000-17000r_g/c$ for $a=0.5$.}
    \label{fig:time_evolution}
\end{figure*}

Figure \ref{fig:time_evolution} shows the time evolution of the accretion rate $\dot M$ in code units (a), magnetic flux $\Phi = 0.5 \int d\theta d \phi \sqrt{-4 \pi g} |B^r|$ though a hemisphere (b), and dimensionless magnetic flux $\phi_{\rm BH}=\Phi/\sqrt{\dot M r_g^2 c}$ (c) measured at $2r_g$ as functions of time, starting from $8000r_g/c$. Here $g$ is the determinant of spherical Kerr-Schild metric. Spins of $0.5$ and $0.98$ are shown by blue and black lines respectively. The time periods chosen for the GR radiative transfer in \S~\ref{sec:imaging} (shown by shaded blue and grey regions for $a=0.5$ and $a=0.98$ respectively) are such that the accretion rate is almost constant and no magnetic flux eruptions occur. We have also performed the same analysis for different quiescent time periods and found no qualitative difference in the obtained results.   The time interval we use to calculate average images is relatively short but we do not analyze the time variability properties of our results so this modest time interval is sufficient for our purposes.

\bibliography{main}{}
\bibliographystyle{aasjournal}

\end{document}